\documentclass[preprint]{aastex}
\input psfig.sty

\def\spose#1{\hbox to 0pt{#1\hss}}
\def\simlt{\mathrel{\spose{\lower 3pt\hbox{$\mathchar"218$}}
     \raise 2.0pt\hbox{$\mathchar"13C$}}}
\def\simgt{\mathrel{\spose{\lower 3pt\hbox{$\mathchar"218$}}
     \raise 2.0pt\hbox{$\mathchar"13E$}}}

\slugcomment{Submitted to AJ}

\shorttitle{The Evolution of Radio Galaxies}
\shortauthors{Brown et al.}

\begin{document}

\title{The Evolution of Radio Galaxies at Intermediate Redshift}

\author{M. J. I. Brown}
\affil{National Optical Astronomy Observatory, P.O. Box 26732, 950 North Cherry Avenue, Tucson, AZ 85726, USA}
\email{mbrown@noao.edu}

\author{R. L. Webster}
\affil{School of Physics, University of Melbourne, Parkville, Victoria 3010, Australia}

\and

\author{B. J. Boyle}
\affil{Anglo-Australian Observatory, P.O. Box 296, Epping, NSW 1710, Australia}

\begin{abstract}
We describe a new estimate of the radio galaxy
$1.4 {\rm GHz}$ luminosity function and its evolution at intermediate 
redshifts ($z \sim 0.4$). Photometric redshifts and color 
selection have been used to select $B_J<23.5$ early-type
galaxies from the Panoramic Deep Fields, a multicolor survey of 
two $25\Box^\circ$ fields. Approximately 230 radio galaxies have then 
been selected by matching early-type galaxies with NVSS radio sources
brighter than $5 {\rm mJy}$. Estimates of the $1.4 {\rm GHz}$ 
luminosity function of radio galaxies measure significant 
evolution over the observed redshift range. For an $\Omega_M=1$ cosmology 
the evolution of the radio power is consistent with luminosity 
evolution where $P(z)\sim P(0)(1+z)^{k_L}$ and $3<k_L<5$. 
The observed evolution is similar to that observed for $UVX$ and X-ray
selected AGN and is consistent with the same physical process 
being responsible for the optical and radio luminosity evolution of 
AGN. 
\end{abstract}


\section{Introduction}

Evolution of radio sources is required to explain the observed 
number of radio sources and the redshift distribution
of radio sources with observed optical counterparts 
\citep{lon66, con84, pea85}. Radio emission from high redshift AGN
is detectable in wide-field radio sky surveys and the optical 
counterparts of some $z>3$ radio-loud QSOs can be detected on 
Schmidt photographic plates. However, most radio-loud AGN
are radio galaxies and at $z \sim 0.5$ their faint optical 
counterparts ($B_J\sim 23$) make it difficult to obtain the large
spectroscopic samples required to measure the evolution of the 
luminosity function.

Several previous samples of radio galaxies with redshifts 
are summarized in Table~\ref{table:rgprev}. To date, most estimates 
of the radio galaxy luminosity function have relied on 
catalogues limited to bright radio flux limits \citep{kue81, wal85, dun89}
or bright optical flux limits \citep{con89, sad89, mac99}.
These catalogues contain relatively few radio galaxies 
(as opposed to radio-loud QSOs) at $z>0.2$, and consequently shed 
little light on the radio galaxy luminosity function and its evolution at 
these redshifts. Samples that have attempted to extend our 
knowledge of the radio galaxy luminosity function 
at $z\simgt 0.2$, by extending the identification of mJy radio 
sources to faint optical magnitudes ($B>22$) have been limited 
to relatively small samples of objects (10-60 galaxies) due to 
the difficultly of obtaining follow-up spectroscopy.  
As a result, estimates of the radio galaxy luminosity function
and evolution obtained from these samples are still subject 
to significant uncertainty.

With photometric redshifts and color selection, it is possible
select and estimate the redshifts of faint early-type galaxies
in multicolor imaging surveys \citep{bro00}. By matching 
early-type galaxies with sources from the radio catalogues, it is 
possible to compile a large uniformly selected catalogue 
of radio galaxies to measure the evolution of the radio galaxy 
luminosity function. As the radio galaxies are selected from deep 
wide-field imaging data, a large catalogue of radio-quiet galaxies 
is also available to measure the environments 
of the radio galaxies \citep{bro01}.

\section{The Panoramic Deep Fields}

The Panoramic Deep Fields \citep{phd} are 
$5^\circ\times 5^\circ$ images of the South Galactic Pole (SGP) 
and UK Schmidt field 855 (F855). The images were produced by 
stacking SuperCOSMOS scans of UK Schmidt photographic
plates in $U$, $B_J$, $R$ and $I$ bands. Object detection, 
instrumental photometry and faint object star-galaxy 
classifications were determined with SExtractor \citep{ber96}. 
Photometric calibration of the data was determined with CCD 
images and published photometry with corrections for dust 
extinction being derived from the dust maps of \cite{sch98}.
The final galaxy catalogues are complete to $B_J\sim 23.5$ and $R\sim 22$.
\cite{phd} provides a detailed description of the data reduction and 
calibration of the Panoramic Deep Fields. 

\begin{figure}
\psfig{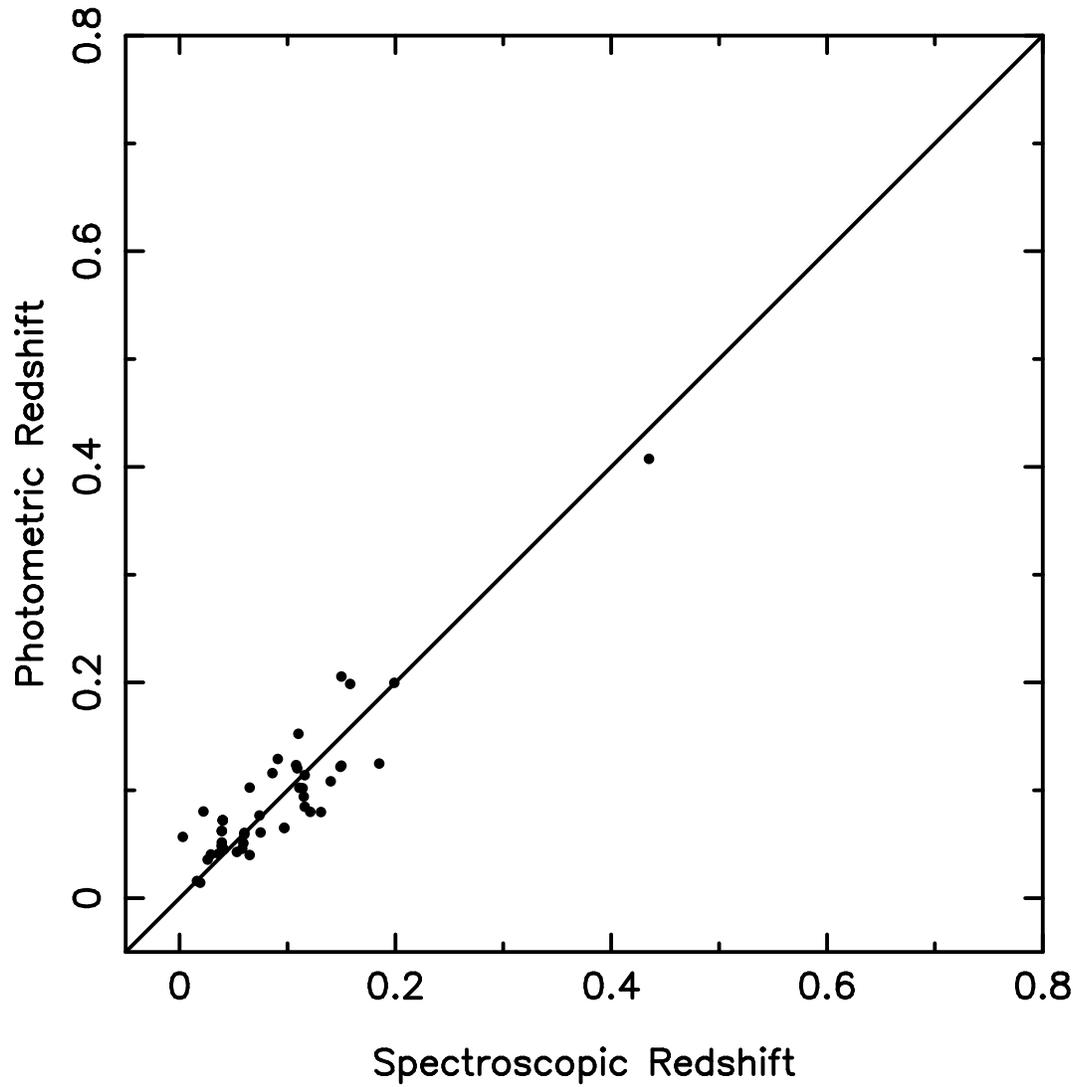}
\figcaption[MJBROWN.fig2.ps]
{A comparison of photometric and spectroscopic redshifts of 
red galaxies which have been matched to the NVSS catalogue. The photometric
redshifts for radio galaxies are $15\%$ larger than the redshifts of 
radio-quiet galaxies with the same multicolor photometry. 
A detailed discussion of the calibration of the photometric redshifts
is provided by \cite{bro00}.
\label{fig:nvssphotoz}}
\end{figure}

Photometric redshifts were calibrated using the polynomial fitting method
of \cite{con95} and $\sim 700$ Panoramic Deep Field galaxies 
with spectroscopic redshifts available from the NED database \citep{bro00}. 
Comparison of the photometric redshifts of radio galaxies with 
spectroscopic redshifts showed that radio galaxy redshifts were 
systematically underestimated by $\sim 15\%$. This is
almost certainly due to radio galaxies being more luminous 
and having slightly bluer colors than other early-type galaxies 
\citep{sad89,gov00}
To correct for this, the photometric redshifts of 
radio galaxies in this paper were increased by $15\%$ and 
Figure~\ref{fig:nvssphotoz} shows there is good agreement between the
corrected photometric and spectroscopic redshifts. 

\section{Selection of Radio Galaxies}

The radio galaxy sample consists of Panoramic Deep Field galaxies 
which are associated with objects in the NVSS source catalogue \citep{con98}. 
$B_J-R$ color and photometric redshifts were used to select  $B_J<23.5$ 
galaxies which are redder than a non-evolving Sbc where the 
Sbc colors were determined with the $k-$corrections of \cite{col80}. 
The optical galaxies were then matched to radio sources
from the NVSS source catalog. 
The selection of early-type galaxies removes contamination 
of the radio galaxy sample by QSOs and starbursts. 
If unified models of radio-loud AGN \citep{bar89} are valid, the removal of
QSOs from the sample removed radio galaxies viewed at certain 
orientations. However, as the galaxy photometric redshifts are 
not usable for QSOs, the removal of QSOs from the sample allowed the
measurement of radio galaxy evolution without spectroscopic redshifts. 

As the evolution of the radio galaxy population was studied, the 
criterion used for matching radio and optical sources could not be a 
function of redshift. High completeness was also required to prevent
the estimate of the luminosity function being dominated by 
corrections for incompleteness. Low contamination of the sample 
was required so estimates of the luminosity function and
radio galaxy environments would not be dominated by galaxies
not associated with radio sources. 

The brightest NVSS source catalogue positions have $1\sigma$ 
error estimates of $\sim 1^{\prime\prime}$ while the 
faintest ($\sim 2.5 {\rm mJy}$) source positions have errors of 
$\simgt 10^{\prime\prime}$ \citep{con98}. 
The sky surface density of $B_J<23.5$ galaxies which meet the color selection 
selection criterion is $\sim 0.5 / \Box^{\prime}$ so 
$\simgt 40\%$ of $\sim 2.5 {\rm mJy}$ sources have an unassociated 
optical galaxy within $3\sigma$ of the radio source position. 
To prevent the catalogue being dominated by spurious matches to 
faint radio sources, the radio flux limit was set to $5 {\rm mJy}$ 
which is slightly brighter than completeness limit of the NVSS catalogue.

The Panoramic Deep Fields contain $\sim 9\times 10^4$ early-type galaxies
and $\sim 10^4$ radio sources making the manual selection 
of radio galaxies prohibitively time consuming.
An obvious automated criterion for selecting optical counterparts to radio 
sources is to match objects within  $\sim 3\sigma$ of the 
radio source centroid. However, as many extragalactic radio 
sources have extended emission, this will be incomplete for 
a significant fraction of radio sources. 
Selection criterion using only the angular separation between radio
and optical sources can select a large fraction of all radio sources. 
However, for a fixed angular scale, the projected distance 
increases with redshift introducing selection effects which are a 
function of redshift. Large radio sources will be excluded at low 
redshift but will be included at high redshift where the angular 
size of the radio sources is less than the selection criterion. 

As the optical counterparts of radio sources are physical associations, 
a selection criterion based on the physical separation of the 
optical and radio sources can be used to select radio galaxies. 
While the NVSS catalogue does not contain redshift information, 
photometric redshifts can be used to estimate the redshifts of 
optical galaxies and the projected distance between the optical 
and radio sources. As the size (in physical coordinates) of 
radio sources does not rapidly evolve between $z\sim 1$ and $z\sim 0$ 
\citep{buc98}, a selection criterion using a fixed projected distance 
should not introduce selection effects which are a function of redshift. 

\begin{figure}
\psfig{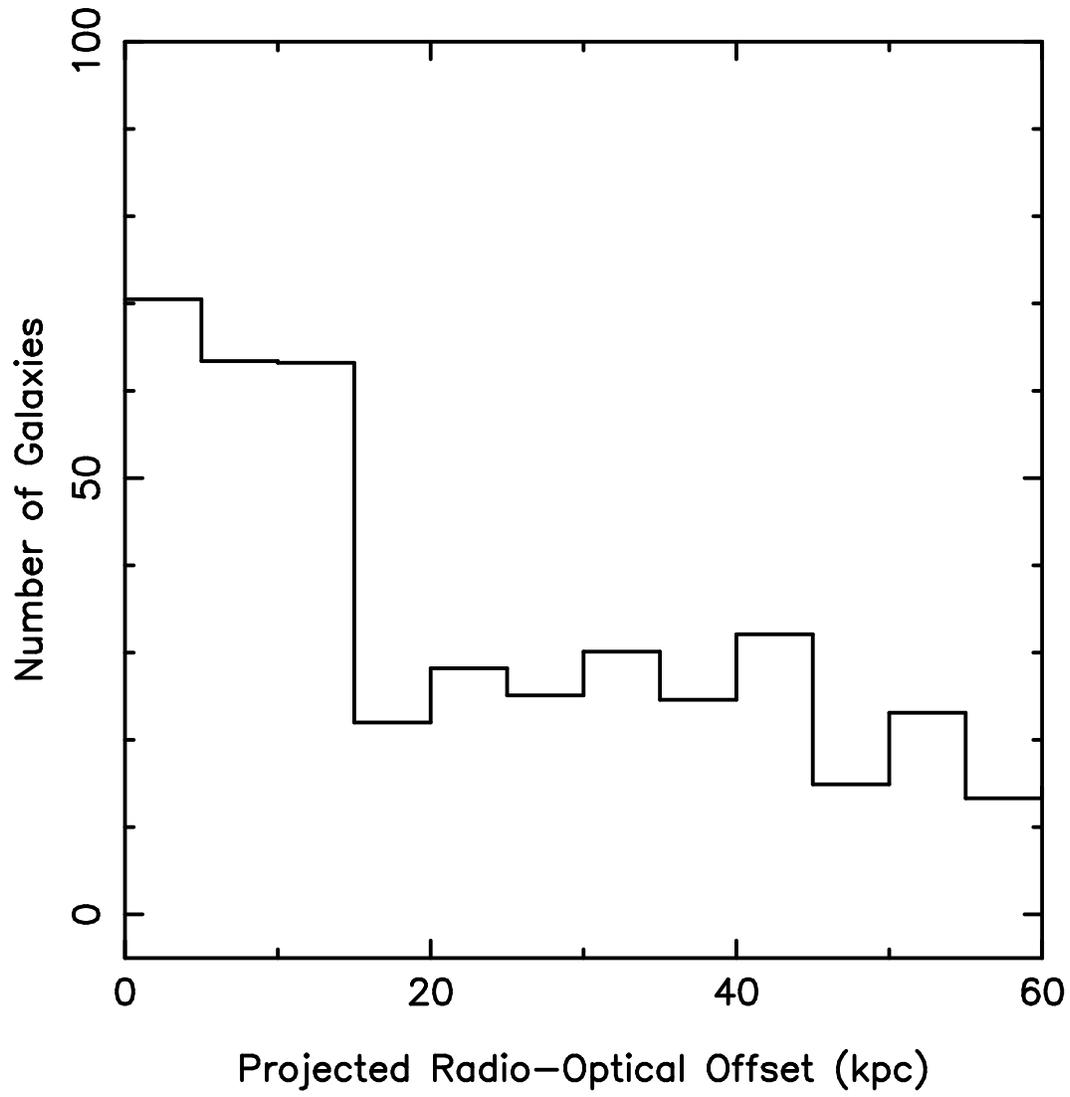}
\figcaption[MJBROWN.fig1.ps]
{The projected distances between radio sources and optical 
galaxies with angular separations $<30^{\prime\prime}$.
\label{fig:rohist}}
\end{figure}

To determine the projected distance used as the selection criterion for
optical counterparts, the distribution of projected distances between 
NVSS sources and optical galaxies within $30^{\prime\prime}$
of NVSS sources was measured and is plotted in 
Figure~\ref{fig:rohist}. The number
of optical counterparts declines rapidly on scales $<20 h^{-1} {\rm kpc}$ 
($h \times 100 {\rm kms}^{-1} {\rm Mpc}^{-1} \equiv H_0$)
after which the distribution has a long tail. Approximately 
$25\%$ of the NVSS sources with early-type galaxies within 
$30^{\prime\prime}$ have more than one early-type counterpart which is 
consistent with much of the tail in Figure~\ref{fig:rohist}
being contamination. To reduce the contamination, optical 
galaxies  $>20  h^{-1} {\rm kpc}$ from NVSS sources were 
excluded from the radio galaxy sample. A complete list of the radio 
galaxy sample is provided in Table~\ref{table:rg_list}. 

The completeness of the selection criterion was estimated 
by applying the $20 h^{-1} {\rm kpc}$ criterion to 
radio galaxies from the Las Campanas Redshift Survey 
identified by \cite{mac99}. Approximately $85\%$ 
of radio galaxies identified by \cite{mac99} were selected
with the $20 h^{-1} {\rm kpc}$ criterion. Fluxes in the 
NVSS source catalogue are estimated by fitting elliptical Gaussians
to the radio emission and for $\sim 5\%$ of radio galaxies where
the fits are poor, the fluxes disagree with \cite{mac99} by $>5\%$.
At $z<0.05$, an increasing fraction of radio galaxies have
radio emission resolved into multiple components by the NVSS 
resulting in $\sim 50\%$ of radio galaxies having source catalogue 
flux estimates which disagree with \cite{mac99}. 

While contamination is low with the $20 h^{-1} {\rm kpc}$ selection criterion, 
it is still present in the radio galaxy catalogue. To prevent these objects
from biasing the estimate of the luminosity function, a control sample 
was constructed by randomly repositioning the radio sources and matching
them to the galaxy catalogue. This process was repeated 10 times
so the control sample size was large enough to not be a significant
source of noise in the estimate of the luminosity function.

Approximately $2\%$ of radio sources with optical counterparts have 
2 optical counterparts resulting in ambiguous redshift 
estimates. As this is a small fraction of the total catalogue, 
rejecting
these objects from the radio galaxy and control samples only
reduced the completeness of the sample by $4\%$ and did not 
significantly bias estimates of the evolution of radio galaxies. The estimates
of the space density and luminosity function parameters were renormalized
to account for the $2\%$ of radio galaxies removed from the catalogue. 
The final catalogue of radio galaxies contains 
230 objects while the control sample (generated with 10 times 
as many radio sources) contains 229 objects. 

\section{Redshift Distribution}

\begin{figure}
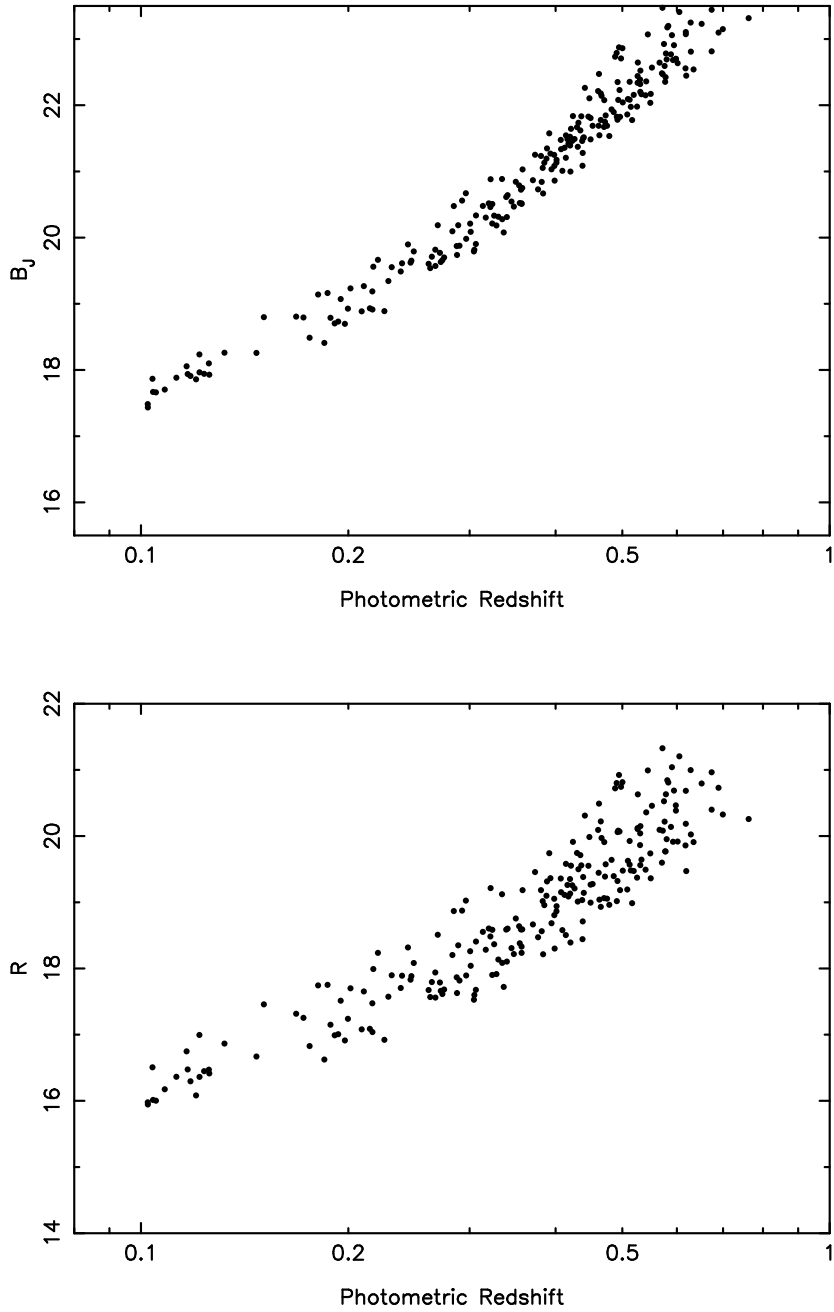

\psfig{file=MJBROWN.fig3a.ps,height=9.0cm,angle=270}
\vspace*{0.25cm}
\psfig{file=MJBROWN.fig3b.ps,height=9.0cm,angle=270}
\figcaption[MJBROWN.fig3a.ps MJBROWN.fig3b.ps] 
{The $B_J$ and $R$ Hubble diagrams of $B_J<23.5$ 
radio-galaxies in the Panoramic Deep Fields.
\label{fig:rghub}}
\end{figure}

\begin{figure}
\psfig{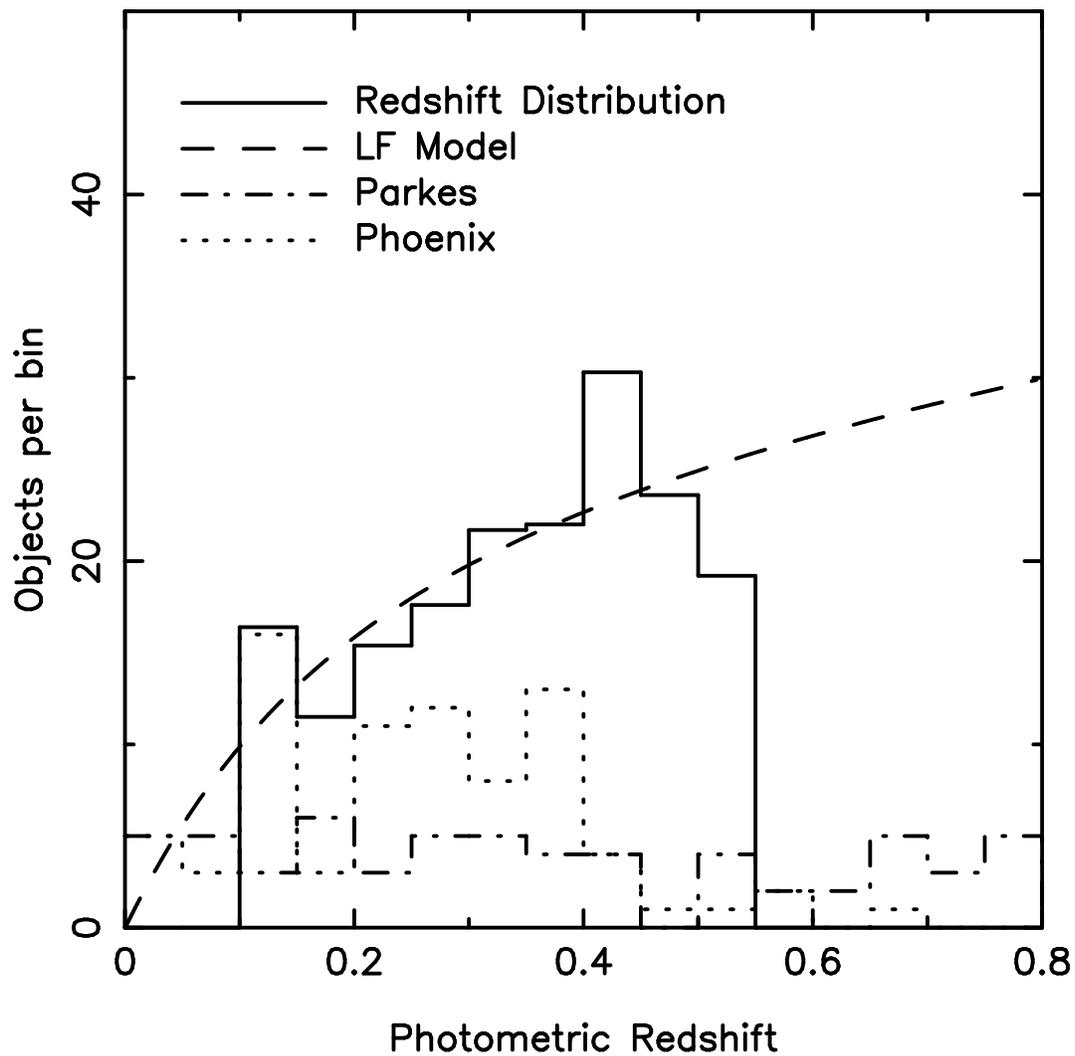}
\figcaption[MJBROWN.fig4.ps]
{The radio galaxy redshift distributions of the Panoramic Deep Fields,
Parkes selected regions\citep{dun89} and Phoenix\citep{hop98}. 
The large size of the Panoramic Deep Fields allows
it to accurately measure the radio galaxy luminosity function 
at intermediate redshifts. 
A model for the Panoramic Deep Fields redshift 
distribution determined with the luminosity function parameters in 
Table~\ref{table:rglfmax} is shown with the dashed line.
\label{fig:rgdndz}}
\end{figure}

The $B_J$ and $R$ Hubble diagrams for the optical 
counterparts of NVSS sources are shown in Figure~\ref{fig:rghub}. 
There is a strong correlation between redshift and magnitude though this
is partially due to the use of photometric redshifts which 
result in objects with the same multicolor photometry having the 
same redshift estimate. The $R$ band Hubble diagram shows evidence
of incompleteness at $z>0.55$ due to the $B_J$ magnitude limit. 
As the apparent magnitude of a $M_{B_J}\sim -20$ elliptical at 
$z\sim 0.55$ is $B_J\sim 23$, this is not unexpected. 
As the incompleteness at $z>0.55$ could significantly
bias estimates of radio galaxy evolution, the sample was limited
to redshifts $z<0.55$. 
 At low redshifts, the errors of the
photometric redshifts are comparable to the redshift estimates resulting
in large errors for the estimates of the radio power. To prevent these objects from 
introducing errors into the estimate of the luminosity function, only the 
196 radio galaxies with $0.10<z<0.55$ were used
to estimate the luminosity function.  The redshift distributions 
of the radio galaxy sample, the Parkes selected regions\citep{dun89} and 
Phoenix\citep{hop98} are shown in Figure~\ref{fig:rgdndz}. 
The size of the Panoramic Deep Fields results in an improved
estimate of the radio galaxy luminosity function at 
$z\sim 0.4$ and the broad redshift distribution has allowed
the evolution of $z<0.55$ radio galaxies to be measured. 

\section{Color Distribution}

\begin{figure}
\psfig{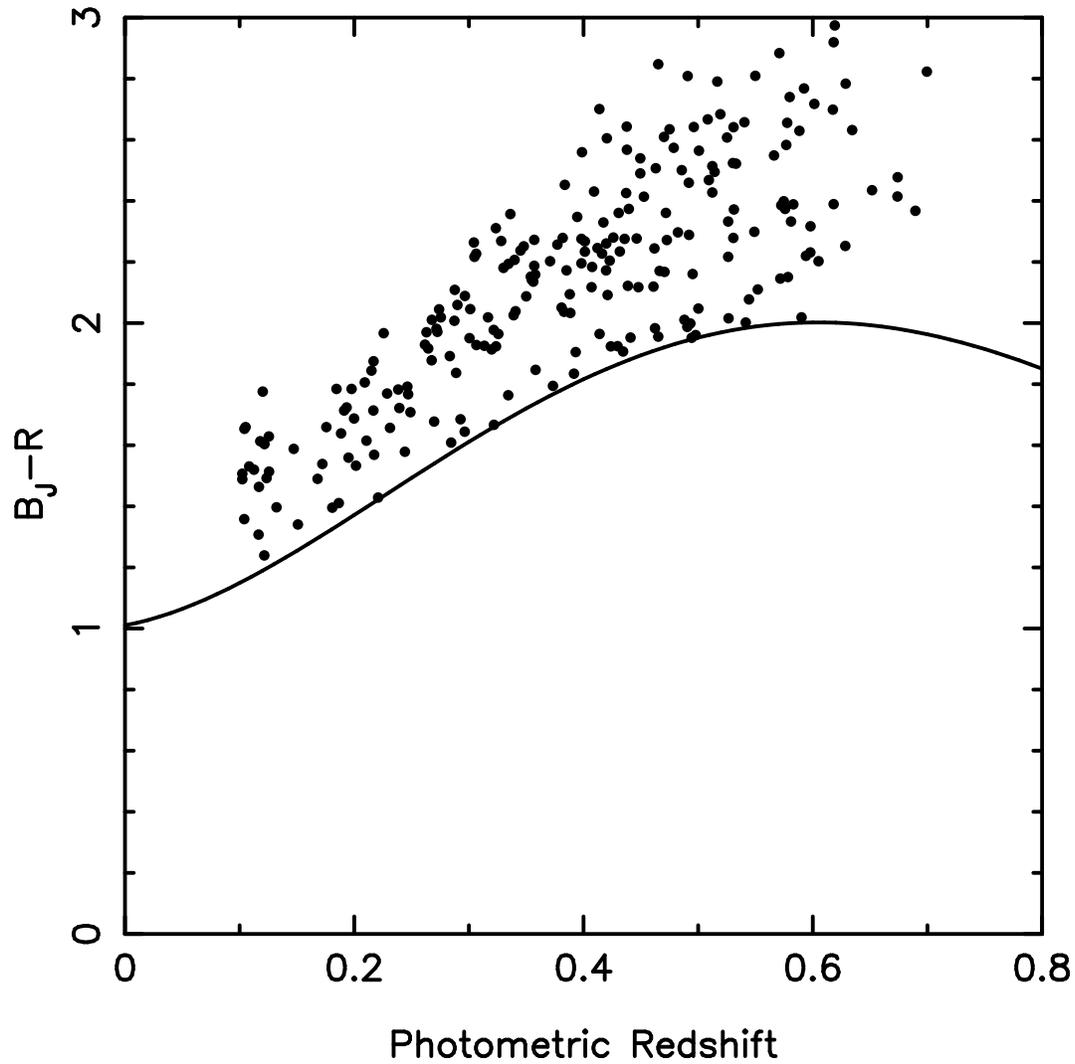}
\figcaption[MJBROWN.fig5.ps]
{The colors of radio galaxies as a function of photometric
redshift. The Sbc color selection criterion is shown with the curved
line. Most radio galaxies have $B_J-R$ colors
significantly redder than the selection criterion.
\label{fig:rgcol}}
\end{figure}

The color selection criterion applied to the radio galaxy candidates
will affect the sample completeness at some level. If a large fraction of 
radio galaxies have colors bluer than the selection criterion, 
significant uncertainty is introduced into estimates of the 
luminosity function and its evolution. 
The colors of galaxies in the radio galaxy sample 
are shown as a function of redshift in Figure~\ref{fig:rgcol}.
While most radio galaxies have red colors, some do have colors 
similar to the selection criterion. However, this could be 
due to contamination which is comprised mostly of blue galaxies 
which dominate the faint galaxy population. 

\begin{figure}
\psfig{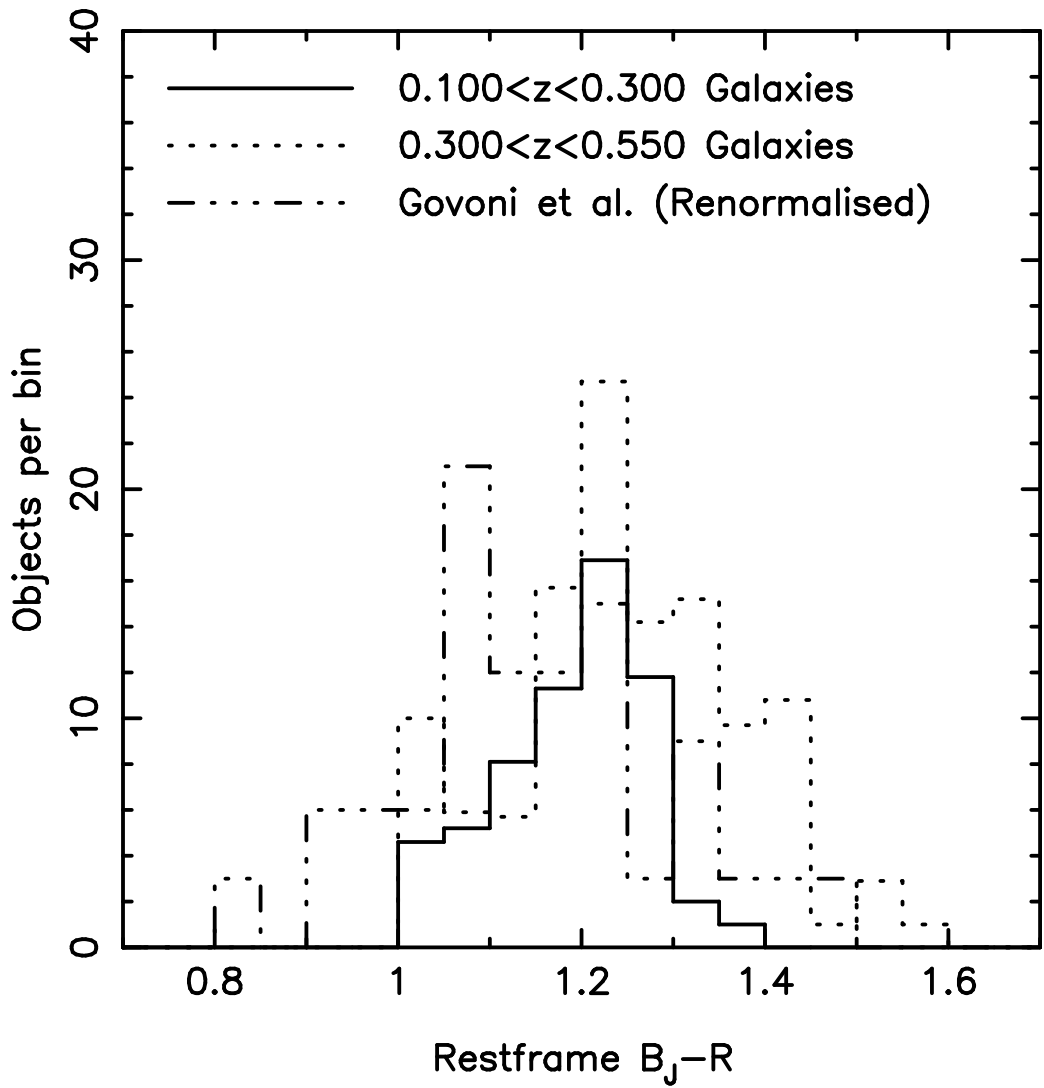}
\figcaption[MJBROWN.fig6.ps]
{The restframe $B_J-R$ colors of the radio galaxy sample and
\cite{gov00}.  
The control sample has been subtracted from the radio galaxy sample 
distribution to remove the affects of contamination. 
The number of objects in the \cite{gov00} has 
been multiplied by 3 to aid comparison of the samples.
\label{fig:restcol}}
\end{figure}

The estimated restframe colors of the radio galaxy sample are plotted
in Figure~\ref{fig:restcol}. Restframe colors were estimated
with linear extrapolations from the E and Sbc $k-$corrections from 
\cite{col80}. The peak of the distribution is 
$\sim 0.2$ magnitudes redder than the $B_J-R=1.01$ selection 
criterion. For comparison, the distribution of radio galaxy 
colors from \cite{gov00} is also shown. If \cite{gov00} sample is 
representative of all radio galaxies, then $\sim 85\%$ of radio 
galaxies were selected by the color selection criterion. However, 
the distributions in Figure~\ref{fig:restcol}
are not identical and a systematic $\delta (B_J-R)=0.1$ 
difference between the \cite{gov00} and Panoramic Deep 
Field colors could alter the completeness by $\sim 10\%$. 
For the remainder of the paper, the incompleteness due to the
color selection criterion is assumed to be $15 \%$. 

\section{The 1.4 GHz Luminosity Function}
\label{sec:rglf}

The restframe radio power of the sources was estimated with
\begin{equation}
{\rm log} P_{1.4} ({\rm WHz}^{-1}) = 
{\rm log} S_{1.4}({\rm mJy}) - (1-\alpha_r){\rm log}(1+z) 
+ 2 {\rm log} D_l ({\rm Mpc}) + 17.185
\end{equation}
where $S_{1.4}$ is the observed $1.4 {\rm GHz}$ radio flux, 
$D_l$ is the luminosity distance and $\alpha_r$ is the radio
spectral index which is assumed to be $0.7$. 
The radio luminosity function was estimated with
\begin{equation}
\Phi(P_{1.4}) = \sum_i \frac{1}{V_{max,i}}
\end{equation}
\citep{sch68}
where $V_{max,i}$ is the maximum comoving volume in which the $i$th source
would be included in the sample. The value of $V_{max, i}$ is given by 
\begin{equation}
V_{max, i} = \Omega \eta_i 
 \frac{c}{H_0} 
\left. \int^{z_{max}}_{z_{min}} \frac{dV_c}{dz} dz \right.
\end{equation}
where $\Omega$ is the survey area, $\eta_i$ is the completeness of the 
survey for sources with the properties of source $i$, $z_{min}$ and 
$z_{max}$ are the minimum and maximum redshifts where source $i$
would be included in the sample and $V_c$ is the comoving volume.
The values of $z_{min}$ and $z_{max}$ were set by the bright 
and faint flux limits (optical and radio) for the sample
and the lower and upper limits for the redshifts over which the luminosity
function is determined. 

\begin{figure}
\psfig{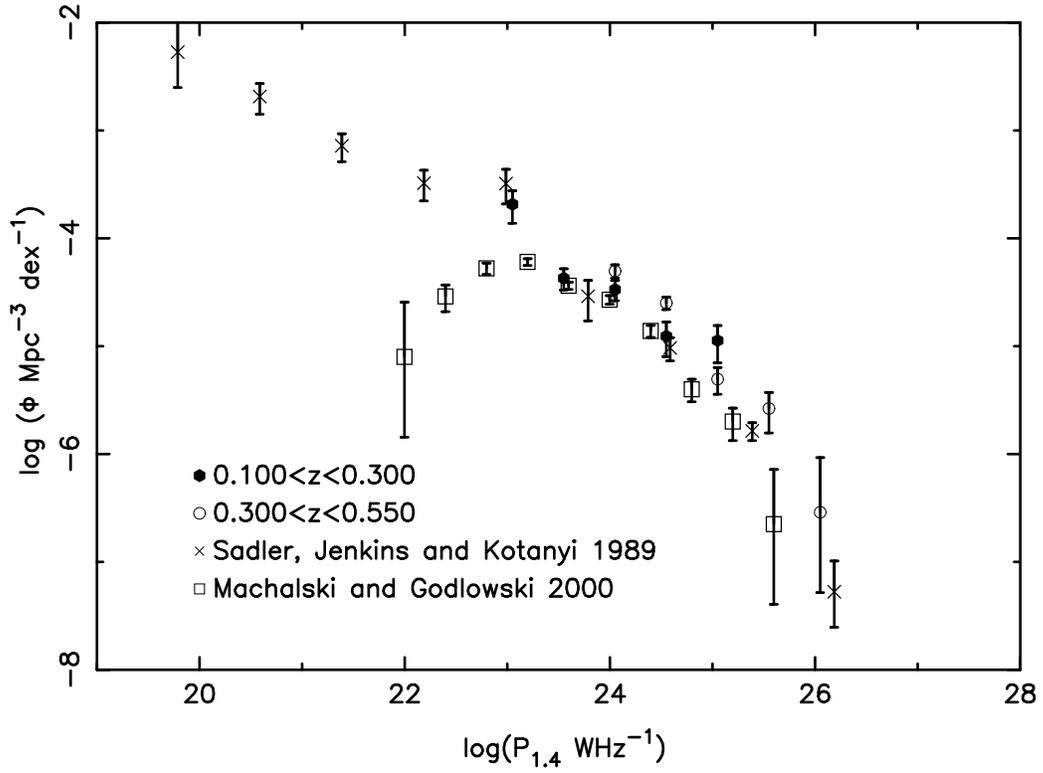}
\figcaption[MJBROWN.fig7.ps]
{The luminosity function of radio galaxies detected in the
Panoramic Deep Fields. Estimates of the luminosity function 
have been corrected for incompleteness. 
For several luminosity bins, the estimate of the space density of 
$0.30<z<0.55$ radio galaxies is significantly higher than the 
estimates for $0.10<z<0.30$ radio galaxies.
\label{fig:rglf}}
\end{figure}

The radio luminosity function of radio galaxies for $0.10<z<0.30$ and 
$0.30<z<0.55$ is shown in Figure~\ref{fig:rglf} and listed in 
Table~\ref{table:rglf}. To be consistent with previous work, the 
estimates of the luminosity function have been determined with 
$\Omega_M=1$ and $H_0=100 {\rm kms}^{-1}{\rm Mpc}^{-1}$. 
The $0.10<z<0.30$ space density is comparable to the low 
redshift samples of \cite{sad89} and \cite{mac00} at 
$P_{1.4}>10^{23} {\rm WHz}^{-1}$. \cite{mac00} may suffer 
incompleteness at $P_{1.4}<10^{23} {\rm WHz}^{-1}$
as their estimate of the space density decreases with decreasing 
radio power. The space density of $0.30<z<0.55$ radio 
galaxies is significantly higher than the low redshift samples 
which is consistent with evolution occurring over the 
observed redshift range. 

To measure the evolution of the radio galaxy luminosity function, the data
was been fitted with a 2 power-law function
\begin{equation}
\Phi(P) d({\rm log} P) = C^* \left[
\left(\frac{P^*(z)}{P}\right)^{\alpha}+
\left(\frac{P^*(z)}{P}\right)^{\beta} \right]^{-1} d({\rm log} P)
\end{equation}
\citep{boy88,dun90}.
The evolution of the luminosity function was assumed to be pure
luminosity evolution where 
\begin{equation}
P^*(z)=P^*(0)(1+z)^{k_L}
\end{equation}
\citep{boy88}. 
The best-fit values for the $C^*$, $P^*(0)$, $\alpha$, 
$\beta$ and $k_L$ were obtained by minimizing 
\begin{equation}
S=-2{\rm ln} L
\end{equation}
\citep{mar83} where $L$ is the likelihood.
The value of $S$ is given by  
\begin{equation}
S= \left. -2 \sum^{N_{rg}}_{i=1} {\rm ln} [ \Phi(P, z_i) ]
+ 2f \sum^{N_c}_{i=1} {\rm ln} [ \Phi(P, z_i) ]
+ 2 \int\int \Phi(P, z)\Omega(P, z) \frac{dV}{dz} dz dP \right.
\end{equation}
where $N_{rg}$ is the number of radio galaxies in the sample, $N_c$ is the number
of control objects and $f$ is the number of radio sources used to produce the radio 
galaxy catalogue divided by the number of radio sources used to generate 
the control catalogue. Error estimates for the measured 
parameters were determined by computing $\Delta S$ for each
parameter in turn while allowing the other values for the parameters
to float \citep{lam76,boy88}.
The errors quoted for the remainder of the 
paper were determined for $\Delta S=1$ which is equivalent to 
$1\sigma$. The goodness-of-fit of the model was tested using the 
2D Kolmogorov-Smirnoff (KS) statistic described by \cite{pea83}. 

\begin{figure}
\psfig{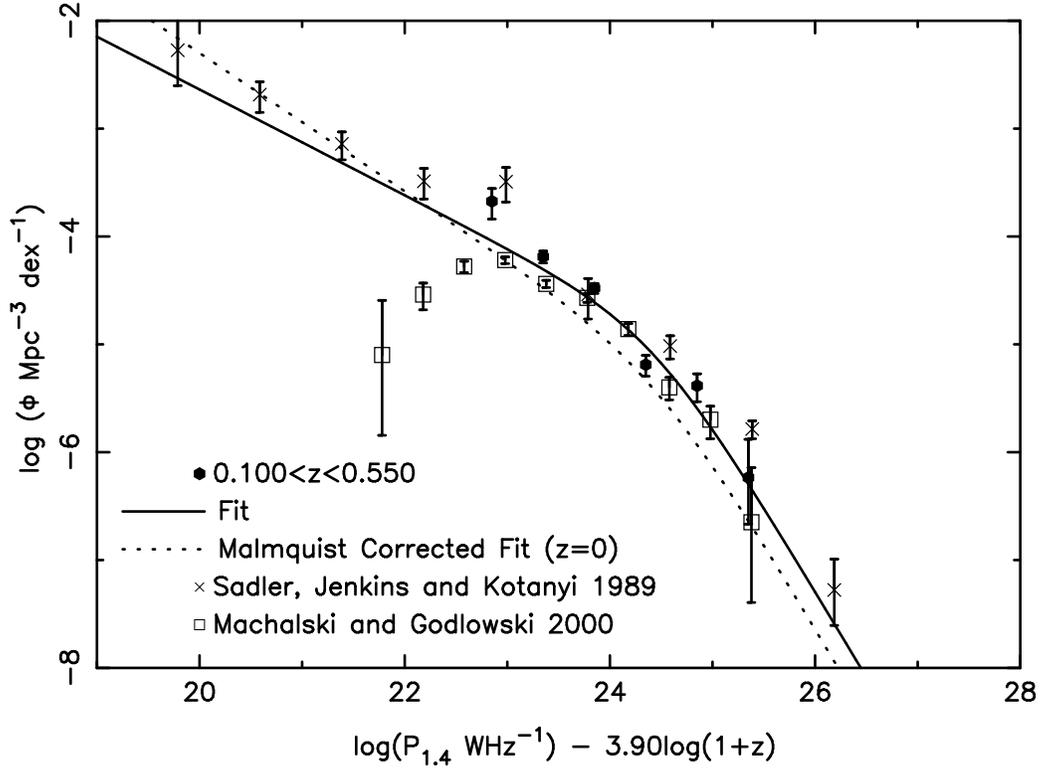}
\caption[MJBROWN.fig8.ps]
{The luminosity function of radio galaxies at the current epoch
where the luminosity at $z\sim 0$ has been estimated by dividing the radio
power by $(1+z)^{3.9}$. The \cite{mac00} data points have been shifted
assuming $z=0.15$. There is reasonable agreement between the 
model of the luminosity function, the data and low redshift surveys 
with the exception of $P<10^{23} {\rm W Hz}^{-1}$ radio galaxies from 
\cite{mac00}. As the size of the Panoramic Deep Fields sample is comparable 
to low redshift samples, the Poisson errors of the luminosity 
function estimates are comparable to \cite{sad89} and \cite{mac00}.
\label{fig:rglfe}}
\end{figure}

The best-fit estimate of the luminosity function is shown in 
Figure~\ref{fig:rglfe} and the estimates of luminosity function 
parameters and the KS probability are listed in Table~\ref{table:rglfmax}. 
To allow comparison with low redshift samples, the radio power of sources
used to estimate the space density have been divided by the estimate
of the luminosity evolution. The luminosity function model is 
comparable to estimates of the luminosity function at low redshift. 
Also, as shown in Figure~\ref{fig:rgdndz}, the model is a good approximation to the 
observed redshift distribution of Panoramic Deep Field radio galaxies. 

The estimates of the luminosity function parameters are affected 
by Malmquist bias due to errors in the radio flux and redshift estimates. 
To correct for this, artificial datasets of radio galaxies
were generated using 2 power-law functions. The fluxes and redshifts 
were then scattered assuming their error distributions can be 
approximated by Gaussians. The distribution of $1\sigma$ values for 
the radio flux error distribution were determined using the error estimates 
provided by the NVSS source catalogue. For the photometric redshifts, 
the errors were estimated by computing the rms of the difference between 
photometric and spectroscopic redshifts as a function of redshift 
for early-type galaxies. 
The parameters which reproduced the observed luminosity function are 
listed in Table~\ref{table:rglfmax}. With the exception 
of $\beta$, most parameters changed by $\simlt 1\sigma$ and as 
shown in Figure~\ref{fig:rglfe}, the
estimate of the luminosity function at $z \sim 0$ is comparable
the original estimate. 

The estimate of the luminosity function and its evolution assumes that the
radio galaxy catalogue is dominated by steep spectrum radio sources 
which is consistent with previous estimates of the luminosity function 
\citep{dun90}. However, if a significant fraction of the radio galaxy 
catalogue are flat-spectrum sources, a smaller value of $\alpha_r$ 
would be more appropriate for estimating the radio powers and the 
luminosities of the $z\sim 0.5$ radio galaxies would be significantly 
decreased. However, as shown in Table~\ref{table:rglfmax}, decreasing the 
value of $\alpha_r$ to $0.5$ only decreases the estimate of 
$k_L$ by $\sim 1 \sigma$.

The estimates of the luminosity evolution
are comparable to previous estimates of the evolution of radio-loud
AGN including QSOs \citep{dun90}. The similarity of 
radio-loud QSO and radio galaxy evolution is consistent with
the unified theory of radio-loud AGN \citep{bar89} where
radio-loud QSOs and radio galaxies are the same class of 
object observed at different orientations. 
The radio evolution of radio galaxies is also similar to the 
evolution of radio-quiet AGN in the optical and X-ray 
\citep{boy88, boy94}, starburst galaxies in the radio \citep{row93} 
and the star formation rate in the UV \citep{lil96}. 

While the similarity of the luminosity function and its evolution
to previous work is not unexpected, it does confirm the ability of the 
techniques used in this work to estimate the radio galaxy luminosity 
function. If the color selection and photometric redshifts are 
used to select radio galaxies from deep 4-m imaging, 
it will be possible to accurately measure the decline of the $z>4$ 
radio-loud AGN luminosity function with redshift and place constraints
on the epoch of formation of AGN. The same techniques will also allow 
wide-field surveys to improve estimates of the $z\sim 0.2$ radio 
galaxy luminosity function, clustering and environments.  

\section{Summary}

The Panoramic Deep Fields and the NVSS have been used to compile a catalogue
of $B_J<23.5$ radio galaxies. 
In this paper we report a new estimate of the radio galaxy 
luminosity function at $z>0.1$ based on this sample.
Radio fluxes, multicolor photometry 
and photometric redshifts have been used to select 230 radio galaxies
over $50 \Box ^\circ$. By matching radio sources to optical 
counterparts with selection criterion that are a function of 
projected distance, radio galaxies are selected with high completeness 
without introducing a strong bias as a function of redshift.
There is significant evolution of the radio galaxy luminosity function 
at $z<0.55$ which can be parameterized
by luminosity evolution where $P(z)\sim P(0)(1+z)^{k_L}$ where
$3<k_L<5$. 
The observed evolution of radio galaxies is consistent with 
the same physical process being responsible for the evolution of 
radio galaxies and radio-quiet AGN. As the estimate of the luminosity 
function does not require large numbers of spectroscopic redshifts, it 
will be possible to apply the same techniques to multicolor surveys where
the time required to obtain large numbers spectroscopic redshifts is
prohibitive. 

\section{Acknowledgments}

The authors wish to thank the SuperCOSMOS unit at Royal Observatory
Edinburgh for providing the digitized scans of UK Schmidt 
photographic plates. The authors also wish to thank Nigel Hambly, Bryn Jones
and Harvey MacGillivray for productive discussions of 
the methods employed to coadd scans of photographic plates.
This research has made use of the NASA/IPAC Extragalactic Database
which is operated by the Jet Propulsion
Laboratory, California Institute of Technology, under contract with the
National Aeronautics and Space Administration. Michael Brown
acknowledges the financial support of an Australian Postgraduate Award.

\newpage

\clearpage

\begin{deluxetable}{lcccccc}
\tablecaption{Several previous samples of radio galaxies (excluding QSOs)\label{table:rgprev}}
\tabletypesize{\scriptsize} 
\tablehead{
Survey & Frequency & Flux 	& Mag. & Spectroscopic	& Photometric \\
       &	   & Limit	& Limit	& Redshifts 	& Redshifts   \\
       & (GHz)	   & (mJy)	& 	& 		& 	      }
\startdata
\cite{kue81} & 5.0 & 1000 & $V \sim 23$	& 118 & -  \\ 
\cite{pea81} & 2.7 & 1500 & $V \sim 22$	& 62  & -   \\ 
\cite{wal85} & 2.7 & 2000 & $V\sim24$	& 107 & 37  \\ 
\cite{wal85} $z>0.3$ & 2.7 & 2000 & $V\sim24$	& 14 & 27  \\ 
\cite{con89} & 1.4 & 0.5  & $B_T=12.0$  & 92 & - \\
\cite{dun89} & 2.7 & 100  & $R\sim 24$	& 36 & 78 \\
\cite{dun89} $z>0.3$  & 2.7 & 100  & $R\sim 24$	& 14 & 73 \\
\cite{sad89} & 5.0 & 0.8  & $B_T=13.8$ 	& 49  & -   \\ 
\cite{row93} & 1.4 & 0.1  & $B\sim 22$  & 8  & 3   \\
\cite{hop98} & 1.4 & 0.2  & $R<21.5$ 	& 63  & -   \\
\cite{sad99} & 1.4 & 2.5  & $B_J=19.4$ 	& 52  & -   \\ 
\cite{mac99} & 1.4 & 2.5  & $R\sim18.5$ & 387 & 162 \\
\\
Panoramic Deep Fields $z>0.1$   & 1.4 & 5.0  & $B_J=23.5$  & -   & 230 \\ 
Panoramic Deep Fields $z>0.3$   & 1.4 & 5.0  & $B_J=23.5$  & -   & 164 \\ 
Panoramic Deep Fields $z>0.5$   & 1.4 & 5.0  & $B_J=23.5$  & -   & 58 \\ 
\enddata
\end{deluxetable}

\begin{deluxetable}{lccccccccc} 
\tablecaption{Radio Galaxy Catalogue\label{table:rg_list}}
\tabletypesize{\scriptsize} 
\tablehead{NVSS Source& RA & Declination & $S_{1.4}$ 
& $B_J$ & $U-B_J$ & $B_J-R$ & $R-I$ & Photo & Optical-NVSS \\
 & (J2000) & (J2000) & (mJy) & & & & & $z$ & Offset}\startdata 
J004407-294751 
& 00~44~07.8 & -29~47~49.7 &   95.7 &  23.24 &      - &   2.35 &      - &  0.581 & $  1.7^{\prime\prime}$ \\ 
J004432-285650 
& 00~44~32.9 & -28~56~48.3 &    8.3 &  22.38 &      - &   2.30 &   1.33 &  0.531 & $  2.2^{\prime\prime}$ \\ 
J004434-295222 
& 00~44~34.0 & -29~52~24.0 &    7.8 &  23.17 &      - &   2.94 &   0.51 &  0.618 & $  1.4^{\prime\prime}$ \\ 
J004443-300606 
& 00~44~43.6 & -30~06~04.6 &    5.6 &  22.00 &      - &   2.32 &   1.10 &  0.482 & $  1.6^{\prime\prime}$ \\ 
J004514-292851 
& 00~45~14.5 & -29~28~51.2 &   16.0 &  22.10 &      - &   2.59 &   0.76 &  0.501 & $  1.7^{\prime\prime}$ \\ 
J004607-282301 
& 00~46~07.7 & -28~22~58.8 &    5.7 &  21.70 &      - &   2.11 &   0.79 &  0.421 & $  2.9^{\prime\prime}$ \\ 
J004633-290137 
& 00~46~34.0 & -29~01~40.1 &    7.2 &  18.76 &   0.96 &   1.80 &   0.61 &  0.198 & $  3.1^{\prime\prime}$ \\ 
J004637-295324 
& 00~46~37.7 & -29~53~31.1 &   37.2 &  20.54 &   0.65 &   1.63 &   0.71 &  0.285 & $  7.4^{\prime\prime}$ \\ 
J004649-300729 
& 00~46~49.6 & -30~07~28.9 &   29.2 &  22.79 &      - &   2.03 &      - &  0.488 & $  2.1^{\prime\prime}$ \\ 
J004658-273304 
& 00~46~58.6 & -27~33~03.5 &   15.0 &  20.04 &      - &   2.11 &   0.63 &  0.297 & $  1.3^{\prime\prime}$ \\ 
J004702-271740 
& 00~47~02.8 & -27~17~42.2 &    7.7 &  21.88 &      - &   2.66 &   0.87 &  0.496 & $  2.5^{\prime\prime}$ \\ 
J004723-273111 
& 00~47~23.0 & -27~31~10.8 &   17.4 &  22.04 &      - &   2.63 &   1.18 &  0.525 & $  0.7^{\prime\prime}$ \\ 
J004725-292157 
& 00~47~24.9 & -29~22~00.3 &    8.1 &  22.87 &      - &   2.80 &   1.25 &  0.629 & $  4.7^{\prime\prime}$ \\ 
J004741-270435 
& 00~47~41.8 & -27~04~34.8 &    7.3 &  19.85 &      - &   2.28 &   0.72 &  0.304 & $  2.0^{\prime\prime}$ \\ 
J004752-284403 
& 00~47~52.1 & -28~44~04.2 &   39.8 &  23.13 &      - &   2.10 &      - &  0.544 & $  1.1^{\prime\prime}$ \\ 
J004808-281054 
& 00~48~08.5 & -28~10~51.8 &   14.3 &  22.62 &      - &   2.72 &   1.53 &  0.618 & $  4.0^{\prime\prime}$ \\ 
J004827-290607 
& 00~48~27.4 & -29~06~03.0 &   15.5 &  18.95 &   0.96 &   1.83 &   0.77 &  0.209 & $  5.1^{\prime\prime}$ \\ 
J004831-294207 
& 00~48~31.5 & -29~42~09.1 &    7.7 &  22.42 &      - &   2.02 &   1.84 &  0.542 & $  3.0^{\prime\prime}$ \\ 
J004836-272938 
& 00~48~36.1 & -27~29~42.8 &    6.9 &  20.58 &   1.33 &   2.16 &   1.17 &  0.355 & $  4.9^{\prime\prime}$ \\ 
J004925-300237 
& 00~49~25.6 & -30~02~37.5 &    8.4 &  19.72 &   0.32 &   1.45 &   1.00 &  0.221 & $  0.3^{\prime\prime}$ \\ 
J004954-282535 
& 00~49~54.9 & -28~25~35.8 &    8.9 &  21.07 &      - &   2.45 &   0.88 &  0.409 & $  1.4^{\prime\prime}$ \\ 
J005049-280409 
& 00~50~49.3 & -28~04~09.4 &   33.8 &  21.89 &      - &   2.30 &   0.29 &  0.436 & $  0.8^{\prime\prime}$ \\ 
J005106-262606 
& 00~51~06.6 & -26~26~01.0 &    5.0 &  22.27 &      - &   2.14 &   0.53 &  0.461 & $  5.3^{\prime\prime}$ \\ 
J005109-300851 
& 00~51~09.6 & -30~08~52.1 &   10.3 &  20.90 &      - &   2.30 &   0.91 &  0.382 & $  1.4^{\prime\prime}$ \\ 
J005116-283144 
& 00~51~15.6 & -28~31~32.2 &   53.8 &  17.49 &   0.75 &   1.51 &   0.76 &  0.102 & $ 14.4^{\prime\prime}$ \\ 
J005116-272049 
& 00~51~16.4 & -27~20~46.7 &   12.9 &  21.31 &      - &   2.22 &   0.75 &  0.398 & $  3.2^{\prime\prime}$ \\ 
J005127-282926 
& 00~51~27.5 & -28~29~23.4 &   77.8 &  18.38 &   0.46 &   1.36 &   1.17 &  0.128 & $  3.0^{\prime\prime}$ \\ 
J005127-282926 
& 00~51~28.1 & -28~29~26.9 &   77.8 &  17.70 &   0.61 &   1.44 &   1.05 &  0.104 & $ 10.7^{\prime\prime}$ \\ 
J005132-282442 
& 00~51~32.6 & -28~24~51.7 &   92.1 &  17.54 &   0.72 &   1.53 &   0.82 &  0.102 & $  8.9^{\prime\prime}$ \\ 
J005209-263054 
& 00~52~09.6 & -26~30~55.5 &   10.1 &  22.16 &      - &   2.14 &   0.43 &  0.448 & $  1.9^{\prime\prime}$ \\ 
J005216-253041 
& 00~52~16.2 & -25~30~43.2 &   11.0 &  21.75 &      - &   2.66 &   0.70 &  0.475 & $  2.8^{\prime\prime}$ \\ 
J005223-284112 
& 00~52~23.9 & -28~41~10.2 &    7.8 &  23.50 &      - &   2.50 &   1.41 &  0.674 & $  3.5^{\prime\prime}$ \\ 
J005302-254755 
& 00~53~02.4 & -25~47~56.5 &   22.0 &  22.21 &      - &   2.68 &   1.10 &  0.540 & $  4.2^{\prime\prime}$ \\ 
J005304-281612 
& 00~53~04.9 & -28~16~13.9 &    9.9 &  23.12 &      - &   2.04 &      - &  0.590 & $  2.8^{\prime\prime}$ \\ 
J005312-293605 
& 00~53~12.7 & -29~36~09.1 &   22.9 &  22.85 &      - &   2.01 &      - &  0.490 & $  3.9^{\prime\prime}$ \\ 
J005321-291448 
& 00~53~21.2 & -29~14~49.3 &   47.5 &  21.09 &   0.03 &   1.87 &   1.06 &  0.358 & $  0.9^{\prime\prime}$ \\ 
J005347-290652 
& 00~53~47.6 & -29~06~49.7 &    6.4 &  22.23 &      - &   2.54 &   1.19 &  0.533 & $  4.8^{\prime\prime}$ \\ 
J005400-265729 
& 00~54~01.0 & -26~57~27.9 &   16.6 &  21.52 &  -0.18 &   2.22 &   0.92 &  0.423 & $  1.9^{\prime\prime}$ \\ 
J005400-282116 
& 00~54~01.1 & -28~21~15.8 &    9.5 &  20.79 &      - &   2.28 &   1.05 &  0.377 & $  2.7^{\prime\prime}$ \\ 
J005422-292720 
& 00~54~22.3 & -29~27~27.4 &   19.9 &  19.40 &   0.63 &   1.79 &   0.53 &  0.229 & $  7.9^{\prime\prime}$ \\ 
J005422-294000 
& 00~54~22.5 & -29~40~02.9 &   18.0 &  22.14 &      - &   2.02 &   1.57 &  0.493 & $  2.6^{\prime\prime}$ \\ 
J005438-253357 
& 00~54~37.8 & -25~33~56.4 &    5.3 &  21.81 &      - &   2.38 &   1.09 &  0.472 & $  3.3^{\prime\prime}$ \\ 
J005440-253946 
& 00~54~40.7 & -25~39~43.6 &    7.0 &  18.32 &   0.47 &   1.42 &   0.84 &  0.132 & $  3.8^{\prime\prime}$ \\ 
J005512-270947 
& 00~55~12.5 & -27~09~47.0 &   11.3 &  19.96 &   0.63 &   1.60 &   0.71 &  0.244 & $  0.1^{\prime\prime}$ \\ 
J005513-275208 
& 00~55~13.3 & -27~52~07.7 &   19.1 &  22.76 &      - &   2.34 &   1.66 &  0.598 & $  4.1^{\prime\prime}$ \\ 
J005515-253900 
& 00~55~15.7 & -25~38~59.8 &    7.8 &  20.70 &      - &   2.06 &   0.85 &  0.341 & $  1.3^{\prime\prime}$ \\ 
J005530-272329 
& 00~55~31.0 & -27~23~30.2 &    6.5 &  21.96 &      - &   2.52 &   0.84 &  0.486 & $  1.2^{\prime\prime}$ \\ 
J005542-275056 
& 00~55~42.2 & -27~50~55.9 &    6.2 &  23.16 &  -1.51 &   2.39 &      - &  0.689 & $  4.6^{\prime\prime}$ \\ 
J005548-283609 
& 00~55~48.7 & -28~36~06.4 &   10.4 &  22.87 &      - &   2.43 &   2.20 &  0.674 & $  3.6^{\prime\prime}$ \\ 
J005608-262507 
& 00~56~08.1 & -26~25~06.0 &  163.7 &  21.87 &      - &   2.56 &   0.02 &  0.450 & $  1.5^{\prime\prime}$ \\ 
J005609-283406 
& 00~56~09.3 & -28~34~02.4 &   48.4 &  18.16 &   0.77 &   1.65 &   0.94 &  0.125 & $  4.7^{\prime\prime}$ \\ 
J005613-285535 
& 00~56~13.7 & -28~55~35.2 &    8.0 &  18.03 &   0.86 &   1.62 &   0.96 &  0.122 & $  1.4^{\prime\prime}$ \\ 
J005630-272949 
& 00~56~30.1 & -27~29~49.9 &   86.1 &  19.61 &   0.78 &   1.68 &   0.78 &  0.231 & $  0.3^{\prime\prime}$ \\ 
J005642-254144 
& 00~56~42.3 & -25~41~44.0 &   68.7 &  23.38 &      - &   3.08 &   1.75 &  0.763 & $  0.4^{\prime\prime}$ \\ 
J005642-264710 
& 00~56~42.4 & -26~47~07.8 &   22.9 &  20.14 &      - &   2.38 &   0.88 &  0.336 & $  3.8^{\prime\prime}$ \\ 
J005647-254337 
& 00~56~48.0 & -25~43~37.7 &    8.6 &  22.58 &      - &   2.39 &   0.89 &  0.531 & $  2.6^{\prime\prime}$ \\ 
J005651-295815 
& 00~56~50.9 & -29~58~13.8 &    9.3 &  18.85 &   0.62 &   1.56 &   0.73 &  0.172 & $  2.8^{\prime\prime}$ \\ 
J005656-254626 
& 00~56~56.4 & -25~46~28.8 &   16.2 &  21.91 &      - &   2.29 &   1.12 &  0.473 & $  3.0^{\prime\prime}$ \\ 
J005700-260929 
& 00~57~00.9 & -26~09~29.0 &    5.3 &  22.50 &      - &   2.35 &   1.01 &  0.526 & $  0.3^{\prime\prime}$ \\ 
J005708-254408 
& 00~57~08.8 & -25~44~10.1 &   12.2 &  21.89 &      - &   2.83 &   0.42 &  0.491 & $  2.1^{\prime\prime}$ \\ 
J005722-290716 
& 00~57~22.9 & -29~07~18.3 &    6.8 &  21.45 &      - &   2.28 &   0.85 &  0.420 & $  4.5^{\prime\prime}$ \\ 
J005725-264830 
& 00~57~25.5 & -26~48~37.8 &    9.5 &  19.65 &   0.74 &   1.48 &   0.61 &  0.217 & $  7.6^{\prime\prime}$ \\ 
J005725-264830 
& 00~57~25.7 & -26~48~25.6 &    9.5 &  18.50 &   0.62 &   1.55 &   0.77 &  0.152 & $  6.0^{\prime\prime}$ \\ 
J005733-295315 
& 00~57~33.4 & -29~53~11.4 &    5.4 &  22.53 &      - &   2.41 &   1.54 &  0.572 & $  5.2^{\prime\prime}$ \\ 
J005742-253316 
& 00~57~42.8 & -25~33~16.1 &  238.4 &  21.61 &      - &   1.99 &   1.11 &  0.414 & $  1.5^{\prime\prime}$ \\ 
J005756-252236 
& 00~57~57.0 & -25~22~37.3 &  301.9 &  18.99 &   0.87 &   1.71 &   0.77 &  0.200 & $  1.2^{\prime\prime}$ \\ 
J005820-251830 
& 00~58~21.0 & -25~18~26.6 &   19.1 &  19.93 &   1.58 &   2.03 &   0.90 &  0.287 & $  3.6^{\prime\prime}$ \\ 
J005823-290428 
& 00~58~23.3 & -29~04~28.5 &   17.7 &  18.99 &   0.89 &   1.86 &   0.70 &  0.215 & $  0.4^{\prime\prime}$ \\ 
J005829-285905 
& 00~58~29.2 & -28~59~06.6 &   30.3 &  18.85 &   0.18 &   1.66 &   0.90 &  0.188 & $  1.7^{\prime\prime}$ \\ 
J005832-292614 
& 00~58~32.7 & -29~26~15.6 &   56.2 &  22.29 &      - &   2.18 &   1.11 &  0.495 & $  1.4^{\prime\prime}$ \\ 
J005833-263224 
& 00~58~33.2 & -26~32~25.7 &   25.1 &  22.99 &      - &   2.42 &   0.93 &  0.575 & $  1.5^{\prime\prime}$ \\ 
J005848-280106 
& 00~58~48.5 & -28~01~07.4 &   19.2 &  19.77 &   0.97 &   1.94 &   0.69 &  0.264 & $  2.8^{\prime\prime}$ \\ 
J005854-292657 
& 00~58~54.9 & -29~26~58.4 &   17.0 &  21.11 &   0.08 &   2.06 &   1.15 &  0.383 & $  2.1^{\prime\prime}$ \\ 
J005908-285949 
& 00~59~09.2 & -28~59~44.2 &  448.2 &  17.92 &   0.97 &   1.80 &   0.74 &  0.120 & $  8.4^{\prime\prime}$ \\ 
J005916-253847 
& 00~59~16.4 & -25~38~49.3 &    6.6 &  20.24 &      - &   2.29 &   0.68 &  0.328 & $  1.6^{\prime\prime}$ \\ 
J005936-252155 
& 00~59~36.9 & -25~21~54.2 &    8.4 &  21.19 &      - &   2.29 &   0.87 &  0.401 & $  1.5^{\prime\prime}$ \\ 
J005945-300925 
& 00~59~45.7 & -30~09~24.9 &    8.9 &  21.60 &      - &   2.59 &   1.11 &  0.479 & $  3.7^{\prime\prime}$ \\ 
J005953-281831 
& 00~59~53.3 & -28~18~28.2 &    5.1 &  20.39 &      - &   1.98 &   1.27 &  0.326 & $  4.3^{\prime\prime}$ \\ 
J010001-291945 
& 01~00~01.1 & -29~19~46.0 &   19.0 &  19.68 &   0.97 &   1.81 &   0.57 &  0.246 & $  2.2^{\prime\prime}$ \\ 
J010004-261114 
& 01~00~04.2 & -26~11~17.3 &   12.6 &  21.30 &      - &   2.01 &   1.00 &  0.386 & $  3.1^{\prime\prime}$ \\ 
J010004-261114 
& 01~00~04.5 & -26~11~11.5 &   12.6 &  22.90 &      - &   2.28 &      - &  0.516 & $  5.5^{\prime\prime}$ \\ 
J010006-283206 
& 01~00~07.0 & -28~32~09.3 &   16.9 &  20.58 &      - &   2.21 &   1.11 &  0.357 & $  4.6^{\prime\prime}$ \\ 
J010010-260651 
& 01~00~10.9 & -26~06~55.0 &    5.3 &  22.65 &      - &   2.39 &   1.44 &  0.576 & $  4.1^{\prime\prime}$ \\ 
J010030-263353 
& 01~00~31.0 & -26~33~53.0 &   98.8 &  22.10 &      - &   2.32 &   1.89 &  0.549 & $  1.4^{\prime\prime}$ \\ 
J010048-274021 
& 01~00~49.0 & -27~40~23.3 &   20.0 &  20.37 &   1.53 &   2.23 &   1.00 &  0.340 & $  1.9^{\prime\prime}$ \\ 
J010102-285840 
& 01~01~02.4 & -28~58~46.9 &   18.7 &  19.80 &   1.10 &   2.13 &   0.83 &  0.288 & $  6.8^{\prime\prime}$ \\ 
J010114-284021 
& 01~01~14.1 & -28~40~17.9 &   18.9 &  20.60 &      - &   2.26 &   0.54 &  0.345 & $  4.3^{\prime\prime}$ \\ 
J010129-272202 
& 01~01~29.9 & -27~22~01.9 &    8.6 &  19.66 &   1.22 &   1.95 &   0.83 &  0.262 & $  0.5^{\prime\prime}$ \\ 
J010132-275444 
& 01~01~32.3 & -27~54~44.1 &   18.4 &  22.83 &  -2.18 &   2.65 &   1.00 &  0.588 & $  0.8^{\prime\prime}$ \\ 
J010139-291502 
& 01~01~39.6 & -29~15~02.1 &   17.8 &  22.41 &      - &   2.45 &      - &  0.512 & $  1.9^{\prime\prime}$ \\ 
J010144-261730 
& 01~01~44.5 & -26~17~29.0 &   11.4 &  22.48 &      - &   2.68 &   1.27 &  0.578 & $  5.3^{\prime\prime}$ \\ 
J010204-294201 
& 01~02~04.1 & -29~42~03.5 &   26.9 &  21.89 &      - &   2.30 &   0.60 &  0.446 & $  5.1^{\prime\prime}$ \\ 
J010212-253733 
& 01~02~13.0 & -25~37~31.3 &    6.3 &  20.85 &      - &   2.17 &   0.56 &  0.354 & $  3.9^{\prime\prime}$ \\ 
J010312-251758 
& 01~03~12.5 & -25~18~01.4 &   16.6 &  21.52 &      - &   2.20 &   0.98 &  0.423 & $  3.1^{\prime\prime}$ \\ 
J010312-251758 
& 01~03~12.7 & -25~17~55.1 &   16.6 &  21.88 &      - &   2.60 &   0.28 &  0.462 & $  4.2^{\prime\prime}$ \\ 
J010320-254037 
& 01~03~19.8 & -25~40~35.3 &    8.2 &  20.16 &   1.38 &   1.91 &   0.53 &  0.283 & $  4.4^{\prime\prime}$ \\ 
J010325-282000 
& 01~03~25.9 & -28~20~00.3 &   12.4 &  19.55 &   1.18 &   1.80 &   0.64 &  0.238 & $  1.0^{\prime\prime}$ \\ 
J010401-300109 
& 01~04~01.5 & -30~01~06.1 &    6.1 &  20.92 &      - &   2.58 &   0.54 &  0.399 & $  3.7^{\prime\prime}$ \\ 
J010407-293616 
& 01~04~07.5 & -29~36~21.1 &    5.8 &  18.00 &   0.82 &   1.51 &   0.88 &  0.124 & $  5.9^{\prime\prime}$ \\ 
J010422-254751 
& 01~04~22.1 & -25~47~51.1 &   61.2 &  22.84 &      - &   2.17 &   1.55 &  0.578 & $  4.6^{\prime\prime}$ \\ 
J010438-300451 
& 01~04~37.9 & -30~04~51.9 &   12.0 &  20.78 &      - &   2.16 &   0.84 &  0.356 & $  3.4^{\prime\prime}$ \\ 
J010448-261248 
& 01~04~48.5 & -26~12~50.0 &    8.2 &  22.45 &      - &   2.54 &      - &  0.530 & $  3.1^{\prime\prime}$ \\ 
J010449-283020 
& 01~04~49.8 & -28~30~24.4 &    8.9 &  21.84 &      - &   2.81 &   1.01 &  0.517 & $  4.7^{\prime\prime}$ \\ 
J010451-251721 
& 01~04~51.6 & -25~17~24.8 &   10.0 &  20.57 &   1.32 &   1.94 &   0.98 &  0.324 & $  5.6^{\prime\prime}$ \\ 
J010453-262220 
& 01~04~53.4 & -26~22~21.0 &   18.7 &  21.55 &      - &   2.25 &   0.65 &  0.416 & $  1.1^{\prime\prime}$ \\ 
J010456-294043 
& 01~04~56.9 & -29~40~43.2 &   30.2 &  19.94 &   1.57 &   2.08 &   0.77 &  0.290 & $  1.2^{\prime\prime}$ \\ 
J010520-252313 
& 01~05~21.0 & -25~23~10.5 &    8.3 &  18.76 &   1.13 &   1.73 &   0.75 &  0.191 & $  3.0^{\prime\prime}$ \\ 
J010525-263141 
& 01~05~25.1 & -26~31~44.9 &    5.0 &  20.95 &      - &   1.78 &   0.94 &  0.334 & $  4.1^{\prime\prime}$ \\ 
J010527-281043 
& 01~05~28.2 & -28~10~42.9 &   10.9 &  19.13 &   1.03 &   1.58 &   0.69 &  0.195 & $  4.6^{\prime\prime}$ \\ 
J010542-295441 
& 01~05~42.5 & -29~54~42.7 &    7.0 &  20.90 &      - &   2.11 &   0.49 &  0.350 & $  1.9^{\prime\prime}$ \\ 
J010545-254751 
& 01~05~45.3 & -25~47~52.6 &   34.5 &  19.96 &      - &   2.25 &   0.68 &  0.306 & $  1.1^{\prime\prime}$ \\ 
J010545-295033 
& 01~05~45.7 & -29~50~32.5 &   12.3 &  21.19 &      - &   2.19 &   0.70 &  0.385 & $  0.7^{\prime\prime}$ \\ 
J010547-285728 
& 01~05~47.8 & -28~57~28.2 &   18.2 &  19.25 &   0.65 &   1.73 &   0.86 &  0.217 & $  1.0^{\prime\prime}$ \\ 
J010607-280935 
& 01~06~07.2 & -28~09~34.4 &   10.9 &  22.92 &  -1.29 &   2.07 &      - &  0.500 & $  1.0^{\prime\prime}$ \\ 
J010610-262431 
& 01~06~10.7 & -26~24~31.0 &   63.9 &  20.34 &      - &   2.21 &   0.95 &  0.335 & $  1.1^{\prime\prime}$ \\ 
J010625-284913 
& 01~06~25.9 & -28~49~16.7 &   12.1 &  21.73 &      - &   2.63 &   0.68 &  0.470 & $  3.9^{\prime\prime}$ \\ 
J010632-261545 
& 01~06~32.4 & -26~15~42.6 &   11.6 &  21.64 &      - &   1.86 &   0.82 &  0.392 & $  4.0^{\prime\prime}$ \\ 
J010633-285055 
& 01~06~33.5 & -28~50~53.6 &   12.0 &  20.52 &   0.84 &   2.00 &   0.84 &  0.322 & $  3.0^{\prime\prime}$ \\ 
J103241+013424 
& 10~32~41.0 & +01~34~24.7 &    9.2 &  22.86 &      - &   2.29 &   1.80 &  0.598 & $  0.6^{\prime\prime}$ \\ 
J103409+021227 
& 10~34~09.1 & +02~12~26.0 &    9.2 &  20.35 &   0.13 &   2.00 &   0.79 &  0.301 & $  2.5^{\prime\prime}$ \\ 
J103414+000521 
& 10~34~14.6 & +00~05~21.4 &  254.8 &  19.43 &   0.57 &   1.49 &   0.73 &  0.181 & $  4.6^{\prime\prime}$ \\ 
J103418-002030 
& 10~34~19.1 & -00~20~32.1 &    5.0 &  18.53 &   0.44 &   1.34 &   0.70 &  0.122 & $  2.9^{\prime\prime}$ \\ 
J103424-021102 
& 10~34~24.4 & -02~11~02.6 &  123.5 &  20.41 &   0.69 &   1.75 &   0.58 &  0.270 & $  0.5^{\prime\prime}$ \\ 
J103431+012356 
& 10~34~31.2 & +01~23~55.7 &   15.9 &  22.52 &      - &   2.28 &   1.36 &  0.526 & $  0.4^{\prime\prime}$ \\ 
J103453+003637 
& 10~34~54.0 & +00~36~37.8 &  139.1 &  22.34 &      - &   2.26 &   0.94 &  0.471 & $  0.5^{\prime\prime}$ \\ 
J103456-001007 
& 10~34~56.9 & -00~10~04.3 &    6.3 &  21.65 &      - &   2.13 &   0.75 &  0.389 & $  3.8^{\prime\prime}$ \\ 
J103514-002432 
& 10~35~15.0 & -00~24~30.8 &   16.9 &  21.76 &      - &   2.53 &   0.78 &  0.437 & $  1.5^{\prime\prime}$ \\ 
J103516+003242 
& 10~35~16.2 & +00~32~41.2 &   17.1 &  22.62 &      - &   2.67 &   1.48 &  0.577 & $  1.2^{\prime\prime}$ \\ 
J103521+012020 
& 10~35~20.9 & +01~20~23.3 &    5.5 &  22.82 &      - &   2.07 &   1.24 &  0.526 & $  4.0^{\prime\prime}$ \\ 
J103525-000145 
& 10~35~25.4 & -00~01~40.8 &    7.8 &  21.79 &      - &   2.38 &   0.81 &  0.426 & $  4.5^{\prime\prime}$ \\ 
J103526-012513 
& 10~35~26.3 & -01~25~13.1 &    7.3 &  21.46 &      - &   2.18 &   0.90 &  0.388 & $  1.3^{\prime\prime}$ \\ 
J103552-004246 
& 10~35~52.0 & -00~42~45.6 &    6.1 &  22.13 &      - &   2.76 &   1.02 &  0.508 & $  0.9^{\prime\prime}$ \\ 
J103609+013721 
& 10~36~09.0 & +01~37~20.7 &   11.0 &  21.51 &      - &   2.30 &   0.78 &  0.412 & $  1.2^{\prime\prime}$ \\ 
J103614-022237 
& 10~36~14.8 & -02~22~36.2 &   28.1 &  20.41 &      - &   1.91 &   0.85 &  0.289 & $  2.2^{\prime\prime}$ \\ 
J103618+011403 
& 10~36~19.2 & +01~14~07.6 &   13.4 &  20.38 &      - &   2.37 &  -0.08 &  0.324 & $  5.7^{\prime\prime}$ \\ 
J103619-001955 
& 10~36~19.8 & -00~19~55.8 &    8.4 &  20.64 &   1.17 &   2.03 &   0.81 &  0.306 & $  0.7^{\prime\prime}$ \\ 
J103623+011118 
& 10~36~23.0 & +01~11~19.7 &    5.8 &  21.96 &      - &   2.52 &   1.21 &  0.492 & $  1.9^{\prime\prime}$ \\ 
J103647-004055 
& 10~36~47.2 & -00~40~56.2 &   12.9 &  19.07 &   0.77 &   1.58 &   0.83 &  0.168 & $  2.4^{\prime\prime}$ \\ 
J103700+014542 
& 10~37~00.5 & +01~45~39.5 &   44.7 &  22.34 &      - &   2.69 &   0.92 &  0.531 & $  3.8^{\prime\prime}$ \\ 
J103705-015607 
& 10~37~05.8 & -01~56~06.1 &   45.5 &  22.02 &      - &   2.93 &  -0.34 &  0.465 & $  1.2^{\prime\prime}$ \\ 
J103737-004111 
& 10~37~36.9 & -00~41~17.5 &   29.4 &  18.17 &   0.71 &   1.70 &   0.74 &  0.118 & $  7.8^{\prime\prime}$ \\ 
J103740-012801 
& 10~37~40.9 & -01~27~59.8 &   81.8 &  21.77 &      - &   2.25 &   0.84 &  0.420 & $  3.6^{\prime\prime}$ \\ 
J103749-000521 
& 10~37~49.2 & -00~05~22.9 &   32.0 &  21.54 &      - &   1.89 &   1.18 &  0.373 & $  2.3^{\prime\prime}$ \\ 
J103828-002932 
& 10~38~29.0 & -00~29~34.1 &    5.3 &  20.90 &      - &   2.53 &   0.84 &  0.384 & $  4.5^{\prime\prime}$ \\ 
J103834+011924 
& 10~38~34.9 & +01~19~23.2 &    6.8 &  20.77 &   1.42 &   2.08 &   0.92 &  0.339 & $  2.3^{\prime\prime}$ \\ 
J103836+011753 
& 10~38~36.8 & +01~18~00.7 &  321.4 &  19.72 &   0.75 &   1.62 &   0.91 &  0.217 & $  6.8^{\prime\prime}$ \\ 
J103843-020738 
& 10~38~43.4 & -02~07~36.2 &    6.6 &  21.61 &      - &   2.44 &   0.93 &  0.431 & $  4.0^{\prime\prime}$ \\ 
J103850-013928 
& 10~38~50.0 & -01~39~25.4 &   10.9 &  20.91 &   0.36 &   1.72 &   0.31 &  0.296 & $  4.2^{\prime\prime}$ \\ 
J103855-015040 
& 10~38~55.6 & -01~50~43.2 &    9.9 &  21.11 &      - &   2.28 &   0.80 &  0.371 & $  3.6^{\prime\prime}$ \\ 
J103908-023917 
& 10~39~08.8 & -02~39~22.8 &   26.5 &  20.77 &   0.03 &   1.76 &   0.35 &  0.293 & $  5.5^{\prime\prime}$ \\ 
J103911+013800 
& 10~39~11.2 & +01~38~01.2 &   24.0 &  21.36 &      - &   2.09 &   0.71 &  0.381 & $  1.4^{\prime\prime}$ \\ 
J103920+010510 
& 10~39~20.4 & +01~05~14.5 &    8.3 &  18.61 &   0.67 &   1.85 &   0.94 &  0.185 & $  6.1^{\prime\prime}$ \\ 
J103938+003046 
& 10~39~38.4 & +00~30~46.4 &    8.4 &  18.20 &   0.63 &   1.55 &   0.90 &  0.117 & $  1.4^{\prime\prime}$ \\ 
J103943+011955 
& 10~39~43.3 & +01~19~55.3 &  126.5 &  21.37 &      - &   2.76 &  -0.43 &  0.414 & $  4.1^{\prime\prime}$ \\ 
J103944-021125 
& 10~39~44.2 & -02~11~28.0 &   29.7 &  22.87 &      - &   2.80 &   1.25 &  0.601 & $  3.0^{\prime\prime}$ \\ 
J103956-024041 
& 10~39~57.0 & -02~40~43.2 &    5.3 &  22.38 &      - &   2.25 &   0.72 &  0.466 & $  1.7^{\prime\prime}$ \\ 
J104006+012304 
& 10~40~05.8 & +01~23~03.4 &    5.3 &  21.85 &      - &   2.30 &   1.24 &  0.462 & $  2.8^{\prime\prime}$ \\ 
J104014-023503 
& 10~40~14.9 & -02~35~04.0 &   17.3 &  22.33 &      - &   2.55 &   1.05 &  0.509 & $  0.8^{\prime\prime}$ \\ 
J104020+010507 
& 10~40~20.3 & +01~05~05.4 &   11.2 &  22.46 &      - &   2.02 &      - &  0.441 & $  3.1^{\prime\prime}$ \\ 
J104037+004223 
& 10~40~37.8 & +00~42~21.9 &   15.9 &  21.80 &      - &   2.59 &   0.99 &  0.463 & $  1.3^{\prime\prime}$ \\ 
J104039+013658 
& 10~40~39.7 & +01~37~03.9 &    5.6 &  19.78 &   1.39 &   2.02 &   1.07 &  0.272 & $  5.6^{\prime\prime}$ \\ 
J104044+020535 
& 10~40~44.9 & +02~05~33.4 &   20.0 &  21.84 &      - &   2.46 &   0.71 &  0.453 & $  2.4^{\prime\prime}$ \\ 
J104105-004334 
& 10~41~05.2 & -00~43~37.4 &   27.0 &  19.00 &   0.63 &   1.41 &   1.12 &  0.151 & $  3.0^{\prime\prime}$ \\ 
J104106+011904 
& 10~41~06.3 & +01~19~03.8 &   11.0 &  22.15 &      - &   2.55 &   1.23 &  0.514 & $  0.5^{\prime\prime}$ \\ 
J104118+001052 
& 10~41~18.1 & +00~10~47.6 &    6.3 &  21.74 &      - &   2.45 &   0.84 &  0.439 & $  4.9^{\prime\prime}$ \\ 
J104121-022303 
& 10~41~21.5 & -02~23~01.1 &    7.9 &  19.18 &   0.70 &   1.97 &   0.70 &  0.217 & $  2.8^{\prime\prime}$ \\ 
J104122+020900 
& 10~41~22.9 & +02~09~02.2 &  250.4 &  21.66 &      - &   2.18 &   1.34 &  0.439 & $  2.2^{\prime\prime}$ \\ 
J104129-003759 
& 10~41~29.1 & -00~38~00.3 &   47.5 &  17.86 &   0.64 &   1.72 &   0.85 &  0.104 & $  0.5^{\prime\prime}$ \\ 
J104131-002736 
& 10~41~31.9 & -00~27~33.4 &   13.7 &  17.86 &   0.76 &   1.73 &   0.94 &  0.105 & $  4.4^{\prime\prime}$ \\ 
J104138-005412 
& 10~41~39.0 & -00~54~16.3 &   20.6 &  22.65 &      - &   3.05 &   1.32 &  0.619 & $  4.2^{\prime\prime}$ \\ 
J104159-020144 
& 10~41~59.6 & -02~01~43.9 &   48.6 &  20.07 &   1.69 &   2.30 &   0.91 &  0.305 & $  3.1^{\prime\prime}$ \\ 
J104208-002116 
& 10~42~08.4 & -00~21~15.4 &   28.9 &  21.08 &  -0.35 &   1.73 &   0.86 &  0.322 & $  1.1^{\prime\prime}$ \\ 
J104212+015213 
& 10~42~12.5 & +01~52~12.1 &   28.5 &  19.74 &   0.82 &   2.07 &   0.84 &  0.268 & $  1.7^{\prime\prime}$ \\ 
J104216+012004 
& 10~42~16.2 & +01~20~02.7 &   23.5 &  21.79 &      - &   1.96 &   1.50 &  0.435 & $  1.4^{\prime\prime}$ \\ 
J104231+003902 
& 10~42~31.5 & +00~39~06.7 &    9.8 &  21.64 &      - &   2.40 &   0.60 &  0.417 & $  4.7^{\prime\prime}$ \\ 
J104241-000947 
& 10~42~41.7 & -00~09~45.9 &   12.5 &  22.38 &      - &   2.02 &   1.06 &  0.465 & $  1.6^{\prime\prime}$ \\ 
J104246-022758 
& 10~42~46.9 & -02~27~57.2 &    7.8 &  22.85 &      - &   2.20 &   1.59 &  0.552 & $  1.7^{\prime\prime}$ \\ 
J104250+003355 
& 10~42~51.0 & +00~33~54.2 &    5.8 &  21.70 &      - &   2.19 &   0.74 &  0.407 & $  3.2^{\prime\prime}$ \\ 
J104301+010648 
& 10~43~01.6 & +01~06~54.6 &   10.9 &  19.72 &   1.08 &   2.03 &   0.91 &  0.263 & $  6.5^{\prime\prime}$ \\ 
J104302+005051 
& 10~43~02.3 & +00~50~47.5 &   56.9 &  19.90 &   1.55 &   2.09 &   0.81 &  0.275 & $  7.5^{\prime\prime}$ \\ 
J104336+011624 
& 10~43~35.8 & +01~16~28.5 &   99.2 &  19.79 &   0.48 &   1.78 &   0.61 &  0.239 & $  5.1^{\prime\prime}$ \\ 
J104335-000736 
& 10~43~35.8 & -00~07~34.0 &   20.8 &  21.22 &      - &   2.41 &   0.71 &  0.395 & $  4.7^{\prime\prime}$ \\ 
J104339-000008 
& 10~43~39.8 & -00~00~14.0 &    5.7 &  19.37 &   0.26 &   1.48 &   0.88 &  0.187 & $  5.9^{\prime\prime}$ \\ 
J104346+010038 
& 10~43~46.0 & +01~00~41.8 &   13.0 &  23.26 &      - &   2.45 &   1.36 &  0.618 & $  3.5^{\prime\prime}$ \\ 
J104347-012404 
& 10~43~48.0 & -01~24~00.2 &    8.7 &  20.96 &      - &   2.23 &   0.79 &  0.358 & $  4.2^{\prime\prime}$ \\ 
J104352+000600 
& 10~43~52.4 & +00~06~03.7 &    7.5 &  22.39 &      - &   2.88 &   0.99 &  0.550 & $  3.4^{\prime\prime}$ \\ 
J104412-023829 
& 10~44~12.6 & -02~38~31.1 &    5.8 &  20.01 &      - &   2.06 &   0.51 &  0.272 & $  1.3^{\prime\prime}$ \\ 
J104414-015147 
& 10~44~14.3 & -01~51~47.7 &    6.8 &  19.87 &   0.99 &   1.84 &   0.75 &  0.247 & $  5.8^{\prime\prime}$ \\ 
J104415-010945 
& 10~44~15.8 & -01~09~46.7 &   15.6 &  21.47 &  -0.47 &   1.97 &      - &  0.393 & $  1.8^{\prime\prime}$ \\ 
J104416+020004 
& 10~44~16.1 & +02~00~05.4 &   26.2 &  23.07 &      - &   2.02 &      - &  0.494 & $  2.2^{\prime\prime}$ \\ 
J104420-011146 
& 10~44~20.3 & -01~11~50.2 &  276.0 &  22.67 &      - &   2.05 &      - &  0.462 & $  3.7^{\prime\prime}$ \\ 
J104455+005654 
& 10~44~55.0 & +00~56~55.3 &   16.6 &  18.06 &   0.56 &   1.58 &   0.86 &  0.113 & $  2.6^{\prime\prime}$ \\ 
J104509-015843 
& 10~45~09.4 & -01~58~43.0 &    9.7 &  22.29 &      - &   2.58 &   1.03 &  0.512 & $  1.4^{\prime\prime}$ \\ 
J104513-004931 
& 10~45~13.4 & -00~49~29.9 &    7.4 &  21.46 &      - &   2.63 &   0.76 &  0.438 & $  3.0^{\prime\prime}$ \\ 
J104513+014557 
& 10~45~13.8 & +01~45~57.1 &   10.1 &  21.28 &      - &   2.71 &   0.91 &  0.438 & $  2.8^{\prime\prime}$ \\ 
J104516-010605 
& 10~45~16.5 & -01~06~07.2 &   12.5 &  20.70 &      - &   1.98 &   0.88 &  0.320 & $  2.1^{\prime\prime}$ \\ 
J104524+001240 
& 10~45~25.0 & +00~12~41.1 &   10.6 &  22.53 &      - &   2.35 &   0.62 &  0.492 & $  0.6^{\prime\prime}$ \\ 
J104537+000119 
& 10~45~37.3 & +00~01~16.3 &   14.7 &  20.69 &      - &   2.34 &   0.92 &  0.357 & $  3.3^{\prime\prime}$ \\ 
J104555+002811 
& 10~45~54.8 & +00~28~10.6 &    9.1 &  21.35 &   0.43 &   2.29 &   0.90 &  0.401 & $  3.5^{\prime\prime}$ \\ 
J104603+014906 
& 10~46~03.8 & +01~49~07.0 &   18.3 &  21.67 &      - &   2.55 &   0.84 &  0.449 & $  1.9^{\prime\prime}$ \\ 
J104603-023014 
& 10~46~03.8 & -02~30~14.4 &   10.6 &  22.66 &      - &   2.95 &   0.75 &  0.571 & $  0.7^{\prime\prime}$ \\ 
J104630-001215 
& 10~46~30.6 & -00~12~12.9 &    6.5 &  21.50 &      - &   2.24 &   0.86 &  0.407 & $  3.5^{\prime\prime}$ \\ 
J104632-011340 
& 10~46~32.4 & -01~13~37.9 &  133.7 &  19.40 &   0.83 &   1.59 &   0.65 &  0.201 & $  2.5^{\prime\prime}$ \\ 
J104633-021714 
& 10~46~33.1 & -02~17~12.8 &   38.1 &  18.67 &   0.83 &   1.72 &   0.74 &  0.176 & $  2.4^{\prime\prime}$ \\ 
J104650-000114 
& 10~46~50.8 & -00~01~09.9 &   35.6 &  22.41 &      - &   2.09 &   0.60 &  0.458 & $  5.5^{\prime\prime}$ \\ 
J104650-000114 
& 10~46~50.9 & -00~01~15.4 &   35.6 &  21.05 &      - &   2.49 &   0.94 &  0.403 & $  0.9^{\prime\prime}$ \\ 
J104658-014729 
& 10~46~58.4 & -01~47~25.7 &   11.5 &  22.73 &      - &   2.69 &   1.85 &  0.634 & $  3.8^{\prime\prime}$ \\ 
J104723-021851 
& 10~47~23.7 & -02~18~49.3 &   30.3 &  20.65 &      - &   1.98 &   0.69 &  0.314 & $  4.0^{\prime\prime}$ \\ 
J104733+001526 
& 10~47~33.5 & +00~15~26.2 &   49.7 &  23.40 &      - &   2.49 &   1.50 &  0.652 & $  1.0^{\prime\prime}$ \\ 
J104738-022255 
& 10~47~38.7 & -02~22~58.9 &  214.4 &  19.97 &   0.32 &   1.77 &   0.66 &  0.249 & $  3.7^{\prime\prime}$ \\ 
J104744+013632 
& 10~47~44.2 & +01~36~37.5 &   28.9 &  23.65 &      - &   2.20 &      - &  0.572 & $  4.7^{\prime\prime}$ \\ 
J104754+012906 
& 10~47~54.8 & +01~29~00.1 &    9.6 &  19.07 &   1.01 &   2.03 &   0.83 &  0.226 & $  6.5^{\prime\prime}$ \\ 
J104801-013016 
& 10~48~01.9 & -01~30~14.8 &    5.4 &  18.04 &   0.26 &   1.42 &   0.81 &  0.104 & $  3.2^{\prime\prime}$ \\ 
J104805+000858 
& 10~48~05.8 & +00~08~58.7 &   47.8 &  22.84 &      - &   2.82 &   0.97 &  0.592 & $  0.7^{\prime\prime}$ \\ 
J104828+003949 
& 10~48~28.7 & +00~39~38.4 &    5.2 &  18.24 &   0.35 &   1.37 &   0.91 &  0.116 & $ 12.0^{\prime\prime}$ \\ 
J104849-011226 
& 10~48~50.0 & -01~12~26.8 &    5.9 &  21.23 &      - &   2.33 &   0.89 &  0.398 & $  1.9^{\prime\prime}$ \\ 
J104905-015747 
& 10~49~05.3 & -01~57~47.1 &    9.0 &  22.00 &      - &   1.98 &   0.88 &  0.424 & $  2.0^{\prime\prime}$ \\ 
J104910-003637 
& 10~49~10.6 & -00~36~40.2 &   18.2 &  23.32 &      - &   2.88 &   1.62 &  0.699 & $  3.5^{\prime\prime}$ \\ 
J104923+000027 
& 10~49~23.1 & +00~00~27.6 &   23.2 &  22.32 &      - &   2.74 &   0.72 &  0.519 & $  0.7^{\prime\prime}$ \\ 
J104926-000343 
& 10~49~26.4 & -00~03~47.9 &    6.7 &  20.48 &      - &   2.24 &   0.79 &  0.330 & $  4.4^{\prime\prime}$ \\ 
J104926+005608 
& 10~49~26.5 & +00~56~09.2 &   33.9 &  18.46 &   0.49 &   1.66 &   0.75 &  0.147 & $  1.1^{\prime\prime}$ \\ 
J104928-022728 
& 10~49~28.6 & -02~27~29.4 &    8.2 &  18.88 &   0.91 &   1.77 &   0.73 &  0.193 & $  1.8^{\prime\prime}$ \\ 
J104933-002743 
& 10~49~33.9 & -00~27~40.9 &   15.3 &  23.58 &      - &   2.26 &   0.97 &  0.605 & $  2.6^{\prime\prime}$ \\ 
J104958-022621 
& 10~49~58.4 & -02~26~23.7 &   14.7 &  22.81 &      - &   2.60 &   0.99 &  0.566 & $  1.9^{\prime\prime}$ \\ 
J105026-020427 
& 10~50~26.9 & -02~04~25.6 &   14.4 &  21.82 &      - &   1.98 &   1.30 &  0.430 & $  3.6^{\prime\prime}$ \\ 
J105036-023618 
& 10~50~36.5 & -02~36~15.9 &   14.5 &  17.86 &   0.70 &   1.58 &   0.92 &  0.108 & $  2.5^{\prime\prime}$ \\ 
J105037-004722 
& 10~50~37.4 & -00~47~24.4 &  125.7 &  23.14 &      - &   2.30 &      - &  0.594 & $  3.9^{\prime\prime}$ \\ 
J105045+004540 
& 10~50~45.5 & +00~45~42.0 &    6.4 &  20.01 &   1.08 &   1.94 &   0.74 &  0.267 & $  4.7^{\prime\prime}$ \\ 
J105047+010738 
& 10~50~47.4 & +01~07~39.5 &   30.4 &  23.44 &      - &   2.32 &   1.44 &  0.628 & $  1.0^{\prime\prime}$ \\ 
J105101-001739 
& 10~51~01.2 & -00~17~39.0 &   16.5 &  20.49 &      - &   2.08 &   0.92 &  0.317 & $  1.9^{\prime\prime}$ \\ 
J105103-021302 
& 10~51~03.7 & -02~12~59.4 &    6.6 &  22.86 &      - &   2.01 &   0.71 &  0.498 & $  3.3^{\prime\prime}$ \\ 
J105111-021317 
& 10~51~11.1 & -02~13~16.5 &   31.6 &  19.42 &   0.30 &   1.67 &   0.75 &  0.211 & $  1.1^{\prime\prime}$ \\ 
J105112-011509 
& 10~51~12.3 & -01~15~10.1 &   23.7 &  20.67 &      - &   2.32 &   0.80 &  0.348 & $  1.1^{\prime\prime}$ \\ 
J105120-014300 
& 10~51~20.6 & -01~42~58.2 &    6.3 &  20.24 &   1.83 &   2.10 &   0.76 &  0.301 & $  3.2^{\prime\prime}$ \\ 
J105126-014404 
& 10~51~26.2 & -01~44~02.7 &   17.2 &  22.84 &      - &   2.79 &   0.82 &  0.580 & $  2.8^{\prime\prime}$ \\ 
J105138-011919 
& 10~51~38.5 & -01~19~20.7 &    5.2 &  18.12 &   0.85 &   1.58 &   0.70 &  0.126 & $  1.5^{\prime\prime}$ \\ 
J105140-022845 
& 10~51~40.6 & -02~28~46.6 &    5.9 &  19.83 &   2.23 &   2.10 &   0.71 &  0.274 & $  5.6^{\prime\prime}$ \\ 
J105150+011351 
& 10~51~50.7 & +01~13~51.4 &   10.1 &  23.39 &      - &   2.45 &   0.67 &  0.583 & $  2.2^{\prime\prime}$ \\ 
J105151+010312 
& 10~51~51.7 & +01~03~16.4 &   16.3 &  21.19 &   0.68 &   2.67 &   0.76 &  0.420 & $  3.9^{\prime\prime}$ \\ 
J105158+005045 
& 10~51~58.9 & +00~50~45.3 &   13.8 &  20.90 &      - &   2.68 &   0.69 &  0.398 & $  0.9^{\prime\prime}$ \\ 
J105158+005045 
& 10~51~59.5 & +00~50~47.6 &   13.8 &  18.84 &   0.89 &   1.66 &   1.29 &  0.155 & $  9.2^{\prime\prime}$ \\ 
J105227-014110 
& 10~52~27.1 & -01~41~08.0 &   23.3 &  21.89 &      - &   2.29 &   0.54 &  0.432 & $  3.7^{\prime\prime}$ \\ 
\enddata 
\end{deluxetable} 
\clearpage

\begin{deluxetable}{cccrr}
\tablecaption{The Radio Galaxy Luminosity Function. Where $k_L=3.9$, the 
radio powers have been divided by the luminosity evolution to 
estimate the luminosity function at $z\sim 0$.\label{table:rglf}}
\tablehead{Redshift Range & $k_L$ & ${\rm Log} (P_{1.4} {\rm W Hz}^{-1})$  & $N_{gal}$ 
& \multicolumn{1}{c}{$\Phi$} $(h^{3} {\rm Mpc}^{-3} {\rm dex}^{-1})$ }
\startdata
$0.10 < z < 0.30$ & 0.0 
& 23.05 & 9 & $2.1\pm^{0.9}_{0.7}\times 10^{-4}$ 	\\
$0.10 < z < 0.30$ & 0.0 
& 23.55 & 20 & $4.2\pm1.0\times 10^{-5}$ 	\\
$0.10 < z < 0.30$ & 0.0 
& 24.05 & 22 & $3.3\pm0.7\times 10^{-5}$ 	\\
$0.10 < z < 0.30$ & 0.0 
& 24.55 & 8  & $1.2\pm^{0.6}_{0.4}\times 10^{-5}$ 	\\
$0.10 < z < 0.30$ & 0.0 
& 25.05 & 7  & $1.1\pm^{0.6}_{0.4}\times 10^{-6}$ \\
\\
$0.30 < z < 0.55$ & 0.0 
& 24.05 & 51 & $5.0\pm{0.7}\times 10^{-5}$ 	\\
$0.30 < z < 0.55$ & 0.0 
& 24.55 & 57 & $2.5\pm{0.3}\times 10^{-5}$ 	\\
$0.30 < z < 0.55$ & 0.0 
& 25.05 & 13 & $5.0\pm^{1.8}_{1.0}\times 10^{-6}$  	\\
$0.30 < z < 0.55$ & 0.0 
& 25.55 & 6 & $2.7\pm^{1.5}_{1.0}\times 10^{-6}$ \\
$0.30 < z < 0.55$ & 0.0 
& 26.05 & 1 & $0.3\pm^{0.6}_{0.2}\times 10^{-6}$ \\
\\
$0.10 < z < 0.55$ & 3.9
& 22.85 & 10  & $2.1\pm^{0.9}_{0.6}\times 10^{-4}$ 	\\
$0.10 < z < 0.55$ & 3.9 
& 23.35 & 63 & $6.5\pm0.8\times 10^{-5}$ 	\\
$0.10 < z < 0.55$ & 3.9 
& 23.85 & 87 & $3.3\pm0.4\times 10^{-5}$ 	\\
$0.10 < z < 0.55$ & 3.9 
& 24.35 & 21 & $6.5\pm1.4\times 10^{-6}$ 	\\
$0.10 < z < 0.55$ & 3.9 
& 24.85 & 12 & $4.1\pm^{1.5}_{0.9}\times 10^{-6}$ 	\\
$0.10 < z < 0.55$ & 3.9 
& 25.35 & 2  & $5.8\pm^{7.4}_{3.7} \times 10^{-7}$ \\
\enddata
\end{deluxetable}

\begin{deluxetable}{lccc}
\tablecaption{The Radio Galaxy Luminosity Function Parameters
\label{table:rglfmax}}
\tablehead{Parameter & Best-Fit Estimate & Malmquist Corrected	& $\alpha_r=0.5$ Estimate }
\startdata
$C^*_{1.4} [ h^3 {\rm Mpc}^{-3} ]$ 
& $(1.5\pm{0.4})\times 10^{-5}  $ 	& $(0.7\pm{0.2})\times 10^{-5}$	& $(1.5\pm{0.4})\times 10^{-5}$ \\
$P^*_{1.4} [ {\rm W} {\rm Hz}^{-1} ]$ 
& $(2.8\pm{0.9})\times 10^{24} $ 	& $(3.2\pm 1.0)\times 10^{24}$ 	& $(3.7\pm{1.2})\times 10^{24} $\\
$\alpha$ 	& $ -1.59\pm{0.10}$ 	& $-1.61\pm{0.10}$ 		& $ -1.71\pm{0.07}$ \\
$\beta$ 	& $-0.49\pm{0.08}$ 	& $-0.64\pm{0.11}$ 		& $ -0.44\pm{0.07}$ \\
$k_L$ 		& $3.9\pm{1.1}$ 	& $4.9\pm{1.4}$ 		& $3.3\pm{1.1}$ \\
$P_{KS}$	& $0.14$		&	-			& $0.20$ \\
\enddata
\end{deluxetable}

\end{document}